\documentclass[pr,twocolumn,superscriptaddress,longbibliography,footnoteadded]{revtex4-1}
\pdfoutput=1

\usepackage[T1]{fontenc}
\usepackage{amssymb}
\usepackage{graphicx}
\usepackage{amsmath}
\usepackage{slashed}
\usepackage{tensor}
\usepackage{hyperref}
\usepackage{epstopdf}
\usepackage{extarrows}
\usepackage{xcolor}

\newcommand{\td}{\text{d}}
\newcommand{\ha}{\hat{a}}

\newcommand{\hc}{\hat{c}}

\newcommand{\Hm}{\mathcal{H}}

\begin{document}
\title{Simulating quantum field theory in curved spacetime with quantum many-body systems}
\author{Run-Qiu Yang}
\affiliation{Center for Joint Quantum Studies and Department of Physics, School of Science, Tianjin University, Yaguan Road 135, Jinnan District, 300350 Tianjin, P.~R.~China}


\author{Hui Liu}
\author{Shining Zhu}
\affiliation{National Laboratory of Solid State Microstructures and School of Physics, Collaborative Innovation Center of Advanced Microstructures, Nanjing University, Nanjing, Jiangsu 210093, China}

\author{Le Luo}
\affiliation{School of Physics and Astronomy, Sun Yat-sen University, 519082, Zhuhai, China}

\author{Rong-Gen Cai}
\email{cairg@itp.ac.cn}
\affiliation{CAS Key Laboratory of Theoretical Physics, Institute of Theoretical Physics, Chinese Academy of Sciences, Beijing 100190, China}

\begin{abstract}
This paper proposes  a new general framework to  build a one-to-one correspondence between quantum field theories in static 1+1 dimensional curved spacetime and quantum many-body systems. We show that a massless scalar field in an arbitrary 2-dimensional static spacetime is always equivalent to a site-dependent bosonic hopping model, while a massless Dirac field  is equivalent to a site-dependent free Hubbard model or a site-dependent isotropic XY model. A possible experimental realization for such a correspondence in trapped ions system is suggested. As applications of the analogue gravity model, we show that they can be used to simulate Hawking radiation of black hole and to study its entanglement.  We also show in the analogue model that black holes are most chaotic systems and the fastest scramblers in nature. We also offer a concrete example about how to get some insights  about  quantum many-body systems from  back hole physics.

\end{abstract}

\maketitle

\section{Introduction}
Quantum field theory in curved spacetime is a semiclassical approximation of quantum gravity theory, where the curved background spacetime is treated classically, while the matter fields in the curved spacetime are quantized. Although  a fully successful quantum gravity theory is still not yet available, such a semiclassical approximation framework has offered us a large amount of interesting new phenomena, such as the Hawking radiation of black hole,  particle production in an expanding universe etc.,  see Refs~\cite{Jacobson:2003vx,Hollands:2014eia} for some review articles. Since in general these phenomena are extremal weak, they are extremely difficult to be observed in the real gravity situations.  
Analogues of black-hole or other phenomena in curved spacetimes in laboratory offer us new perspectives on quantum effects  in curved spacetimes, which might help us deeply understand  the nature of gravity. Following original work of Unruh~\cite{PhysRevLett.46.1351,Unruh:1994je}, which  studies Hawking radiation in sonic analog of a black hole, large amount of systems have been proposed and explored, such as surface wave in water flows~\cite{PhysRevLett.106.021302}, Bose-Einstein condensate (BEC)~\cite{PhysRevLett.85.4643,Steinhauer2016,MuozdeNova2019}, optic systems~\cite{Sheng2013,Bekenstein2017,PhysRevLett.122.010404,Sheng2018,PhysRevLett.120.243901}, ultracold atoms in optical lattices~\cite{Pedernales:2017eue} and so on.
See Refs.~\cite{Barcelo:2005fc,9783319002668,Barcel2018} for  reviews and references therein.

In spite of  impressive progresses have been made  in theoretically and experimentally  on various analogue gravity systems, it is still interesting to seek some analogue models which are more ``pure'' in theory and more easily controlled in experiment. In condensed matter physics, there exist  three basic  quantum many-body models, the hopping model, Hubbard model and isotropic XY model~\cite{Tasaki-1998}, which are of wide applications in many fields. In this paper we find that these models are also of interesting applications in quantum field theory in curved spacetime.  The hopping rate in natural materials is constant, there are no enough motivations for physicists in  condensed matter physics and materials to study the ``site-dependent''  hopping cases.  Here we do see that the ``site-dependent''  hopping  cases occur in the simulation of quantum field theories in curved spacetime by quantum
many-body systems.

So far most of analogue gravity models have  focused on the simulation of Hawking radiation or other types of spontaneous particle creation, such as Unruh effect, particle creation in the  universe, dynamical Casimir effect and so on (see, e.g.,  Refs.~\cite{Hu:2018psq,Trautmann:2016cpg,SchuTzhold:2013fga}).
Let us notice that over the past decades, a remarkable progress in gravity and relevant fields is the proposal of AdS/CFT
correspondence~\cite{Maldacena:1997re,Witten:1998qj,Gubser:1998bc}, which says that a quantum gravity in the anti-de Sitter (AdS) spacetime
is dual to a  conformal field theory (CFT) living in the AdS boundary.  The connection between geometry in the bulk and  entanglement entropy  in the
boundary is also suggested in~\cite{Ryu:2006bv,Nishioka:2009un}.  Recently based on the AdS/CFT correspondence, quantum
scrambling has been suggested as a powerful tool for
characterizing chaos in black holes~\cite{Hayden:2007cs,Shenker2014}, and
Refs.~\cite{Sekino:2008he,Shenker2014,Maldacena2016} conjectured
that  black hole has the  fastest scrambling and is a most quantum
chaotic system in nature.

One of remarkable features of the AdS/CFT correspondence is the strong/weak duality: a weak gravity theory in the AdS  bulk is equivalent to a strong coupled
CFT in the AdS boundary.  Although there exist many pieces of evidence to show the correspondence is true,  it is extremely difficult, if it is not impossible, to prove
the AdS/CFT correspondence.  The analogue gravity models provide the possibility to test experimentally the AdS/CFT correspondence.

In this paper we will show that there exists a one-to-one correspondence  between  quantum field theories in an arbitrary two dimensional spacetime and  site-dependent bosonic hopping model, free Hubbard model or isotropic XY model in quantum many-body  systems. As some applications of our analogue gravity model, we will study Hawking radiation of
black hole and its entanglement, and show that black holes are most chaotic systems and the fastest scramblers in nature, predictions of the AdS/CFT correspondence. We also will use a concrete example to show how to use picture of back hole physics to learn something about quantum many-body systems.



\section{Quantum fields in curved spacetime}
We consider a 2-dimensional background spacetime  with  signature $(+,-)$. In the static case, the metric can always be given in the Schwarzschild coordinates $\{t,x\}$ as,
\begin{equation}\label{metric2d}
  \td s^2=f(x)\td t^2-f(x)^{-1}\td x^2\,.
\end{equation}
In the most cases, we are interested in  the static black hole spacetime with a single non-degenerated horizon, i.e., $f(x)>0$ for $x>x_h$ and there is only a point at $x=x_h$ such that $f(x_h)=0$ but
\begin{equation}\label{defgH}
  g_h=\frac12f'(x_h)>0\,.
\end{equation}
$g_h$ is the surface gravity of the horizon, which gives  the Hawking temperature $T_H=g_h/(2\pi)$ of the black hole. The metric~\eqref{metric2d} in the coordinates $\{t,x\}$ is singular at the horizon. To overcome this shortage, we can define an infalling Eddington-Finkelstein coordinate by the coordinates transformation,
$$t\rightarrow v,~~s.t.,~v=t+\int f(x)^{-1} \td x\,.$$
The metric~\eqref{metric2d} in the infalling Eddington-Finkelstein coordinates $\{v,x\}$ becomes
\begin{equation}\label{metric2c}
  \td s^2=f\td v^2-2\td v\td x\,.
\end{equation}
In this case the metric has no longer the coordinate singularity at the horizon.

%
%

Let us first consider a scalar field in the 2-dimensional curved spacetime. The Klein-Gorden equation of a complex scalar field in the metric~\eqref{metric2c} reads
%
\begin{equation}\label{Lagrphi}
  m^2\phi-2\partial_v\partial_x\phi-f'\partial_x\phi-f\partial_x^2\phi=0\,.
\end{equation}
By introducing  the variable $\varphi$
\begin{equation}\label{defvarphi}
  m\varphi=2\partial_v\phi+f\partial_x\phi\,,
\end{equation}
Eq.~\eqref{Lagrphi} can be rewritten into two coupled 1st order equations
\begin{equation}\label{firstscalar}
\partial_v\phi=-\frac{f}2\partial_x\phi+\frac{m\varphi}2,~~\partial_x\varphi=m\phi\,.
\end{equation}
Now make variable transformation $\phi=w\sqrt{f}$ and we can rewrite the above equations into the  following forms\footnote{This equation is singular at horizon due to $f(x_h)=0$. A discussion about this point can be found in our appendix~\ref{app1}. }
\begin{equation}\label{firstscalar2}
\partial_vw=-\frac{f}2\partial_xw-\frac{f'}4w+\frac{m\varphi}{2\sqrt{f}},~~\partial_x\varphi=mw\sqrt{f}\,.
\end{equation}
%
In the massless limit $m\rightarrow0$,
the above two equations decouple and there is only one independent evolutional equation,
\begin{equation}\label{indepnulleq2}
  \partial_vw=-\frac14[\partial_x(fw)+f\partial_xw]\,.
\end{equation}
%

A similar result can also be obtained for Dirac field. The Dirac equation with the general vielbein ${e^\mu}_a$ and metric $g_{\mu\nu}$ can be written as~\cite{PhysRevD.43.3948,Pedernales:2017eue}
\begin{equation}\label{Diracgeneral}
  i\gamma^a{e^\mu}_a\partial_\mu\psi+\frac{i}2\gamma^a\frac1{\sqrt{-g}}\partial_\mu(\sqrt{-g}{e^\mu}_a)\psi-m\psi=0\,.
\end{equation}
Here $g$ is the determinate of metric $g_{\mu\nu}$. The $\gamma$-matrices in the two-dimensional case are chosen such that $\gamma^a=(\sigma_z, i\sigma_y)$. Choose the vielbein to be
\begin{equation}\label{vielbein1}
  {e^\mu}_a=\left[\begin{matrix}
  -1,&1\\
  -\frac{f}2+\frac12,&\frac{f}2+\frac12
  \end{matrix}\right]\,
  \nonumber
\end{equation}
%
and take the decomposition
$$\psi=\frac1{\sqrt{2}}\left[\begin{matrix}u+w\\u-w\end{matrix}\right]$$
into account, we find that there are two independent equations
\begin{equation}\label{indepnulleq1}
  \partial_vw=-\frac{f}2\partial_xw-\frac{f'}4w+\frac{i}2m u,~~~\partial_xu=-imw\,.
\end{equation}
In the massless limit $m\rightarrow0$, 
there is only one evolutional equation remained, which is the same as Eq.~\eqref{indepnulleq2}.

\section{Map into quantum many-body systems}
\subsection{Theory model}
Now let us discretize the system. The spatial position is
discretized as $x=x_n=nd$ with $n\in \mathbb{N}$ and $d\ll\lambda_0$, where $\lambda_0$ is the effective average wavelength in the system.
The functions in the fixed  spacetime  are then transformed into discrete forms as follows,
$$f_{n}=f(nd),~~w_n(v)=w(v,x_n).$$
The spatial derivatives in Eq.~\eqref{indepnulleq2} are approximated by central differences.
Upon a variable transformation $w_n=(-i)^ne^{-i\mu v}\tilde{w}_n$, Eq.~\eqref{indepnulleq2} can be rewritten into the following form
%
\begin{equation}\label{finiteDeq2}
  i\frac{\td}{\td v}\tilde{w}_{n}=-\kappa_n\tilde{w}_{n-1}-\kappa_{n+1}\tilde{w}_{n+1}-\mu \tilde{w}_n\,.
\end{equation}
with
\begin{equation}\label{defkappan}
  \kappa_n=\frac{f_n+f_{n-1}}{8d}\approx\frac{f[(n-1/2)d]}{4d}\,.
\end{equation}
Here $\mu$ is an arbitrary constant. We will see later that it can be interpreted as the chemical potential in quantum many-body systems. Due to the discretization, discrete form is a well approximation for continuous fields if fields are slowly varying, i.e., Eq.~\eqref{finiteDeq2} is valid in the low energy limit.
%

Now let us quantize these fields themselves. This can be done  by promoting field $\tilde{w}_n$ into operator. For bosonic field, we use the replacement $\tilde{w}_n\rightarrow\ha_n/\sqrt{d}$ and introduce bosonic commutators such that
$$[\ha_n,\ha^\dagger_m]=\delta_{nm},~~[\ha_n,\ha_m]=[\ha^\dagger_n,\ha^\dagger_m]=0\,.$$
The evolutional equation for the field operator then reads,
\begin{equation}\label{finiteDeq3}
  i\frac{\td}{\td v}\ha_{n}=-\kappa_n\ha_{n-1}-\kappa_{n+1}\ha_{n+1}-\mu\ha_n\,.
\end{equation}
%
Considering the evolutional equation in Heisenberg picture $i\partial_v\ha_{n}= [\ha_n,\Hm ]$, Eq.~\eqref{finiteDeq3} implies a following Hamiltonian
\begin{equation}\label{deaHm}
  \Hm=\sum_{n}\left[-\kappa_n\left(\ha_n^\dagger\ha_{n-1}+\ha_{n-1}^\dagger\ha_{n}\right)-\mu\ha_n^\dagger\ha_n\right].
\end{equation}
%
This Hamiltonian describes  a bosonic hopping model and can be treated as a
limit case of a certain of different well-studied quantum systems.
For example, in condensed matter systems, it is the Bose-Hubbard
model~\cite{PhysRev.129.959,PhysRevB.34.3136,PhysRevB.37.325,PhysRevB.40.546}
with site-dependent hopping amplitude and  zero on-site
self-interaction.

For the Dirac field, we can do the similar thing.  Take the replacement $\tilde{w}_n\rightarrow\hc_n/\sqrt{d}$ and introduce anti-commutators such that
$$\{\hc_n^\dagger,\hc_m\}=\delta_{nm},~~\{\hc_n,\hc_m\}=\{\hc^\dagger_n,\hc^\dagger_m\}=0\,,$$
we can obtain a following Hamiltonian form
\begin{equation}\label{deaHm2}
  \Hm=\sum_{n}\left[-\kappa_n\left(\hc_n^\dagger\hc_{n-1}+\hc_{n-1}^\dagger\hc_{n}\right)-\mu\hc_n^\dagger\hc_n\right]\,.
\end{equation}
This is just the free Hubbard model with site-dependent hopping.
This model has been widely studied and can be realized in various different platforms,
see Refs~\cite{Hensgens2017,Tarruell2018,Salfi2016} , for instance. 

The Hamiltonian~\eqref{deaHm2} can also be rewritten into other
well-studied model in condensed matter physics: the isotropic XY
model~\cite{PhysRevA.2.1075,PhysRevA.3.786}. To do that, let us
introduce the following operators according to Jordan-Wigner
transformation
\begin{equation*}
\sigma_n^+=\exp\left[i\pi\sum_{j=1}^{n-1}\hc^\dagger_j\hc_j\right]\hc_n^\dagger,~\sigma_n^-=\exp\left[-i\pi\sum_{j=1}^{n-1}\hc^\dagger_j\hc_j\right]\hc_n
\end{equation*}
and $\sigma_n^z=1-2\hc_n^\dagger\hc_n$ with the
periodic/anti-periodic boundary condition. Upon a constant,
Hamiltonian~\eqref{deaHm2} can be rewritten as
\begin{equation}\label{deaHm3}
  \Hm=\sum_{n}\left[-\kappa_n\left(\sigma_n^+\sigma_{n-1}^-+\sigma_{n-1}^+\sigma_{n}\right)+\frac12\mu\sigma_n^z\right]\,.
\end{equation}
Now introduce the Pauli matrices
\begin{equation}\label{defsigamxyz}
\sigma_n^x=\sigma_n^++\sigma_n^-,~~\sigma_n^y=-i(\sigma_n^+-\sigma_n^-)\,
\end{equation}
then the  above Hamiltonian reads,
\begin{equation}\label{deaHm4}
\Hm=\frac12\sum_{n}\left[-\kappa_n\left(\sigma_n^x\sigma_{n-1}^x+\sigma_{n}^y\sigma_{n-1}^y\right)+\mu\sigma_n^z\right]\,.
\end{equation}
This is nothing, but the isotropic XY model with site-dependent hopping.

\subsection{Experimental simulation}

The Bose-Hubbard model in Eq.~\eqref{deaHm} can be realized in
laboratory with various systems for implementing quantum simulation,
such as optical lattices, superconducting qubits, and trapped ions etc..
Here we just concentrate to a simple case,  which
consists of a linear chain of ions in a linear Paul trap. In a
linear trap, ions are arranged in a Coulomb chain. Assuming $x$ as
one of the transverse directions and $z$ the trap axis, the
Hamiltonian of the chain with $N$ ions is $H=V_0+V_{C}+V_{L}$, where
\begin{equation}
V_0=\frac{1}{2}m\sum_{i=1}^{N}(\omega_{x}^2x_{i}^2+\omega_y^2y_i^2+\omega_z^2z
_i^2)
\end{equation}
\begin{equation}
V_{C}=\sum_{i>j}^{N}\frac{e^2}{\sqrt{(z_i-z_j)^2+(x_i-x_j)^2+(y_i-y_j)^2}}
\end{equation}
\begin{equation}
V_{L}=\sum_{j>i}^{N}t_{i,j}(a_i^\dagger a_j+a_ia_j^\dagger)
\end{equation}
where $\omega_{\alpha},\alpha=x,y,z$ are the trapping frequencies in
each direction, $V_{C}$ is the Coulomb energy, while $V_{L}$ is the
coupling between different axial modes, and $t_{i,j}$ are the hopping
energies that are induced by a pair of Raman laser. For a linear
trap $\omega_{x,y}\gg\omega_z$, the ions form a chain along the $z$
axis and occupy equilibrium positions. Phonons in the $z$ direction
can be described approximately by~\cite{PRL93-263602}
\begin{equation}
H=\sum_{i=1}^{N}\omega_x a_i^\dagger
a_i+\sum_{j>i}^{N}t_{i,j}(a_i^\dagger a_j+a_ia_j^\dagger),
\end{equation}
Note that $t_{i,j}$ can be precisely adjusted to
site(mode)-dependent by varying the phase and the detuning of the
Raman beams. By this scheme, we use the phonon modes of trapped ions
to realize the Bose-Hubbard model with zero on-site energy. To
simulate n-site Hubbard modes, we need to trap N ions in a linear
trap and use N-1 pairs of laser to drive photon transitions between
the N axial modes.

\section{Applications in black hole physics}

\subsection{Hawking radiation and its entanglements}

In this section, we will use the above quantum many-body model to study quantum aspects of black holes in gravity. Let us consider the bosonic hopping model as an example.
For convenience in numerical computations,  let us specify the function $f(x)=\alpha\tanh x$ and $d=0.1$.
In this case, there is a horizon at $x=x_h=0$ with the Hawking temperature $T_H=\alpha/(4\pi)$. It is worth noting that $\kappa_n\neq0$ at the horizon though $f(x)$ is zero at the horizon.
Without lose of generality, we set $\mu=0$ as the total particle number is conserved.

To study the black hole evaporation, we set a particle in the inner region of the black hole by initial state $|\Psi(0)\rangle=|e_{n_0}\rangle$
and choose $n_0=-2/d$ as an example. It describes an initial particle which is localized at the $n_0$-th site. 
Based on the picture of ``pair creation'' in Hawking radiation, ``particle-antiparticle pairs'' can be created around the horizon. The antiparticle (negative energy) falls into the black hole and annihilates with this particle inside the black hole, the particle outside the horizon is materialized and escapes into infinity. Note that the pair creation/annihilation is a virtual processe, and the really materialized result is that the original particle inside the black hole disappears but an identical particle appears outside the horizon. This leads to an equivalent picture to understand Hawking radiation via quantum tunneling: the particle inside the horizon escapes to outside by quantum tunneling. According to Refs.~\cite{Damour:1976jd,Parikh:1999mf,Arzano:2005rs},  neglecting the back reaction of the radiation, the probability of finding this particle outside the horizon and its energy should obey the following blackbody spectrum,
\begin{equation}\label{HawkingPE1}
  P(E)\propto e^{-E/T_H}\,.
\end{equation}
%



%
\begin{figure}
  \centering
  \includegraphics[width=0.22\textwidth]{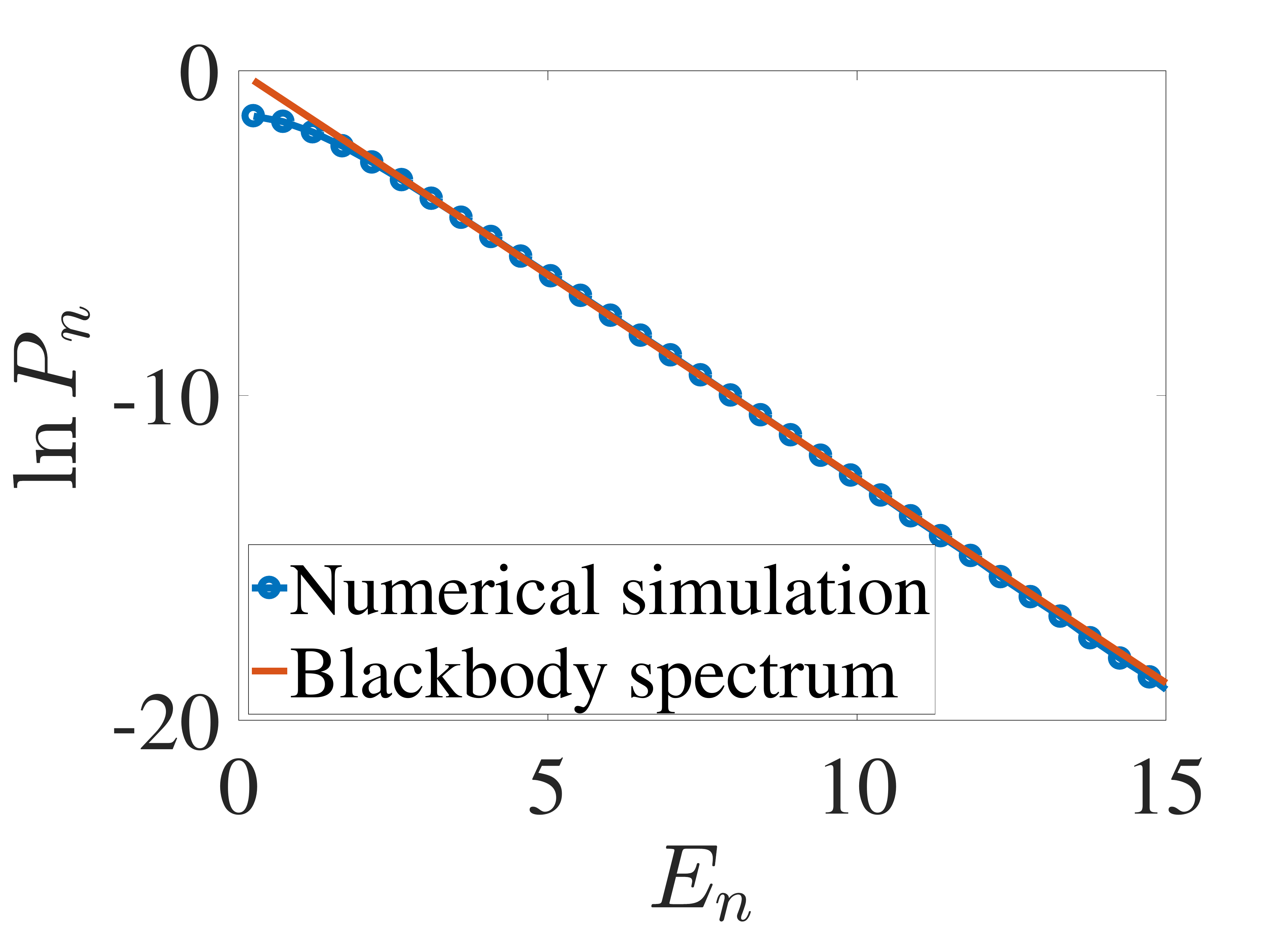}
  \includegraphics[width=0.22\textwidth]{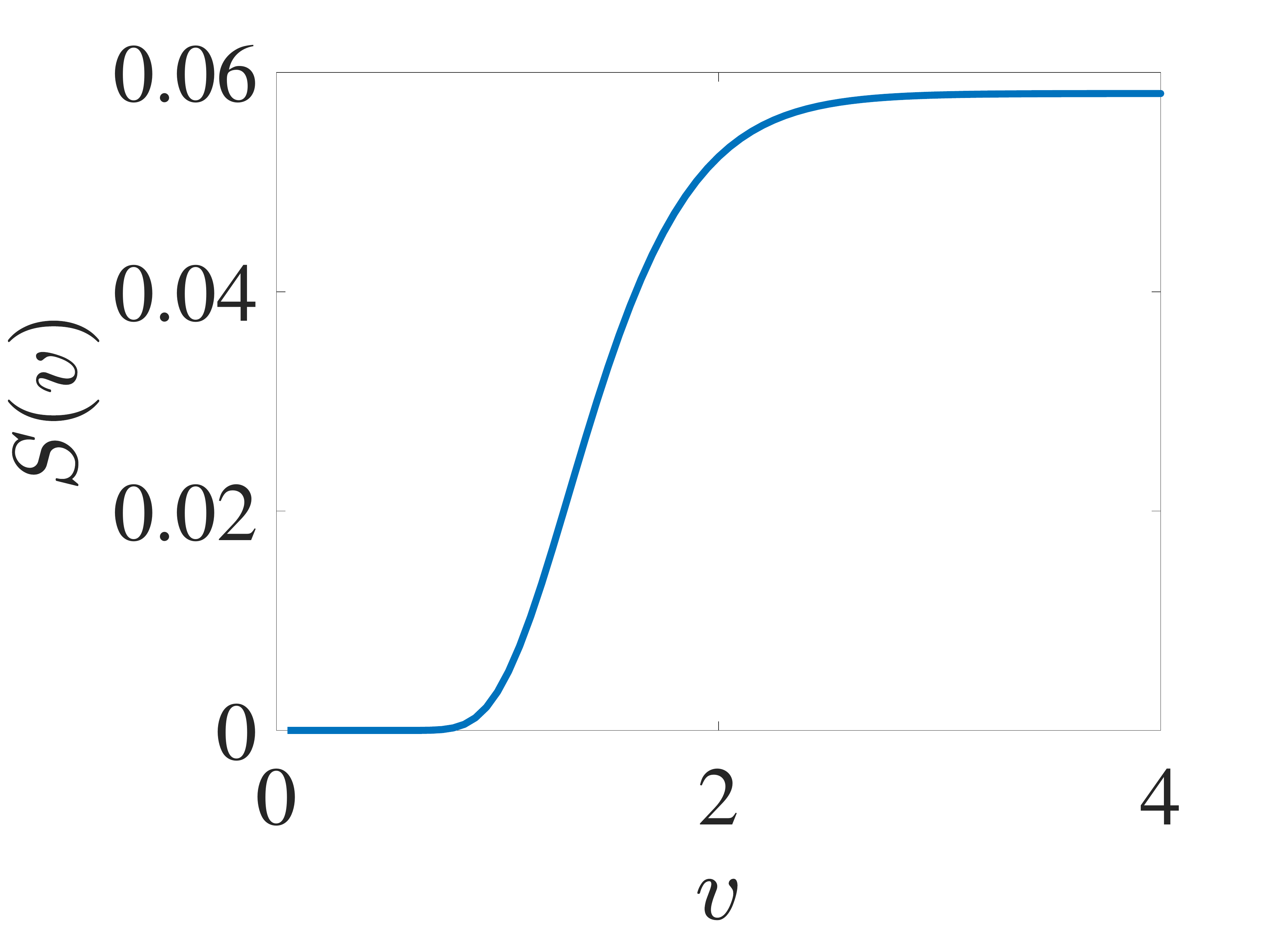}
  \caption{(a): numerical simulation on the probability of finding a particle with energy $E_n$ in the outer region at early time $v=4\lesssim\mathcal{O}(2dL/\alpha)$. (b): the evolution of entanglement entropy between inner region and outer region. The blackbody spectrum is given by Eq.~\eqref{HawkingPE1}. The low energy limit requires that $E_n\ll\mathcal{O}(\alpha/d)$.
  }\label{quantumStPn}
\end{figure}
In Fig.~\ref{quantumStPn} we show the numerical results about the probability of finding a particle of energy $E_n$ in the outer region. Here $E_n$ is the positive eigenvalue of Hamiltonian for the outer subregion. The evolutional time is chosen so that the radiation does not touch the cut-off boundary. For the details of numerical calculation, one may refer to the supplemental materials. We see that numerical results show that $P(E)$ satisfies the blackbody spectrum Eq.~\eqref{HawkingPE1} approximately with the temperature $T=T_H$. Note that the numerical results for smaller energy deviate from the blackbody spectrum~\eqref{HawkingPE1}, because our finite size cut-off cannot cover the low energy region with $E\lesssim\mathcal{O}(2\pi\alpha/(Ld))$ and so leads to the deviation. In addition, we also compute the entanglement entropy between the inner region and outer region, which is given by $S(v)=-\text{Tr}[\rho(v)\ln\rho(v)]$ and the reduced density matrix for outer region is given by $\rho(v)=\text{Tr}_{\text{inner}}(|\Psi(v)\rangle\langle|\Psi(v)|)$. It shows that entanglement between the inner and outer regions  increases  during the Hawking radiation. Because there is only one particle in the black hole, the evaporation will stop in a short time and so the entanglement entropy saturates.

\subsection{Quantum chaos and fastest scrambling}

In this subsection, let us exhibit how to use our analogue model to study some new features of  quantum field theory  in curved spacetime: quantum chaos and fastest scrambling of black holes, appearing from the AdS/CFT correspondence.  To supply an asymptotic AdS$_2$ black hole background, we consider  $f(x)=x^2(1-x_h/x)$ as an example.


To describe the quantum chaos, it was proposed recently that the ``out-time order correlation'' (OTOC) may serve as a useful characteristic of quantum-chaotic behavior. For two local operators $\hat{W}(t)$ and $\hat{V}(t)$ in the Heisenberg picture, their OTOC  is typically defined as
\begin{equation}\label{deotoc}
  C(t):=-\langle[\hat{W}(t), \hat{V}(0)]\rangle\,.
\end{equation}
Here $\hat{W}(0)$ and $\hat{V}(0)$ can be same or different, $\langle\cdot\rangle$ stands for average in an initial state.
Ref.~\cite{Maldacena2016} shows that, with a few general assumptions on the underlying field model and in thermal equilibrium state, the growth of a general OTOC $C(t)$ satisfies
\begin{equation}\label{otocbd1}
  C(t)\propto e^{\lambda_Lt}\,,
\end{equation}
where $\lambda_L$ is the Lyapunov exponent and satisfies following ``chaos bound'',
\begin{equation}\label{otocbd2}
  \lambda_L\leq2\pi T\,.
\end{equation}
Here $T$ is the temperature of the system. The exponential growth~\eqref{otocbd1} will be broken after the ``scrambling time''
\begin{equation}\label{scrambtime}
  t_*\geq\frac{1}{2\pi T}\ln N_f^2,~~~N_f^2\gg1\,.
\end{equation}
Here $N_f$ stands for the degrees of freedom of the system. It is conjectured in~\cite{Sekino:2008he,Shenker2014,Maldacena2016} that  black hole is a most chaotic system and has the fastest scrambling, i.e., it saturates the bounds~\eqref{otocbd2} and \eqref{scrambtime}.

Now let us employ  our model to check if it can exhibit the exponential growth of OTOC and gives us a positive Lyapunov exponent. As an example, we numerically study the following OTOC,
\begin{equation}\label{otoc1}
  C(v):=-\text{Tr}(\rho[\hat{N}_{n_0}(v),\hat{N}_{n_0}]^2)\,.
\end{equation}
Here $\hat{N}_{n_0}$ is a local operator associated to  particle number operator at the $n_0$-th site,
\begin{equation}\label{defN01}
  \hat{N}_{n_0}=\frac{d}{l_0}\sum_{n=-L}^{L}\ha_n^\dagger\ha_ne^{-d^2(n-n_0)^2/l_0^2}\,
\end{equation}
Here $l_0$ has the length scale and stands for the width of distribution of $\hat{N}_{n_0}$.
The time evolutional operator $\hat{N}_n$ is given by Heisenberg picture
$\hat{N}_n(v)=\exp(-i\Hm v)\hat{N}_n\exp(i\Hm v)\,.$
The reason we use the Eq.~\eqref{defN01} to define the local operator $\hat{N}_{n_0}$ rather than $\hat{N}_{n_0}=\ha_{n_0}^\dagger\ha_{n_0}$ is that Eq.~\eqref{defN01} is a well-defined smooth local operator in the continuous limit $d\rightarrow0$. Instead, $\hat{N}_{n_0}=\ha_{n_0}^\dagger\ha_{n_0}$ will become a $\delta$-like function in continuous, which is singular. The initial state is a thermal state with the temperature same as the temperature of black hole
\begin{equation}\label{initiotoc10}
  \rho=\frac1{Z}\sum_{E_{\text{out}}}e^{-\beta E_{\text{out}}} \left |E_{\text{out}}\rangle\langle E_{\text{out}}\right |.
\end{equation}
Here $Z$ is the normalized factor which insures Tr$(\rho)=1$ and the summation contains all the positive energy modes of outside Hamiltonian $\mathcal{H}_{\text{out}}$ (as the negative modes are assumed to fall into black hole). $\mathcal{H}_{\text{out}}$ is obtained by only extracting the sites outside the horizon in  Eqs.~\eqref{deaHm2}, \eqref{deaHm3} and \eqref{deaHm4}.

\begin{figure}
\centering
\includegraphics[width=0.4\textwidth]{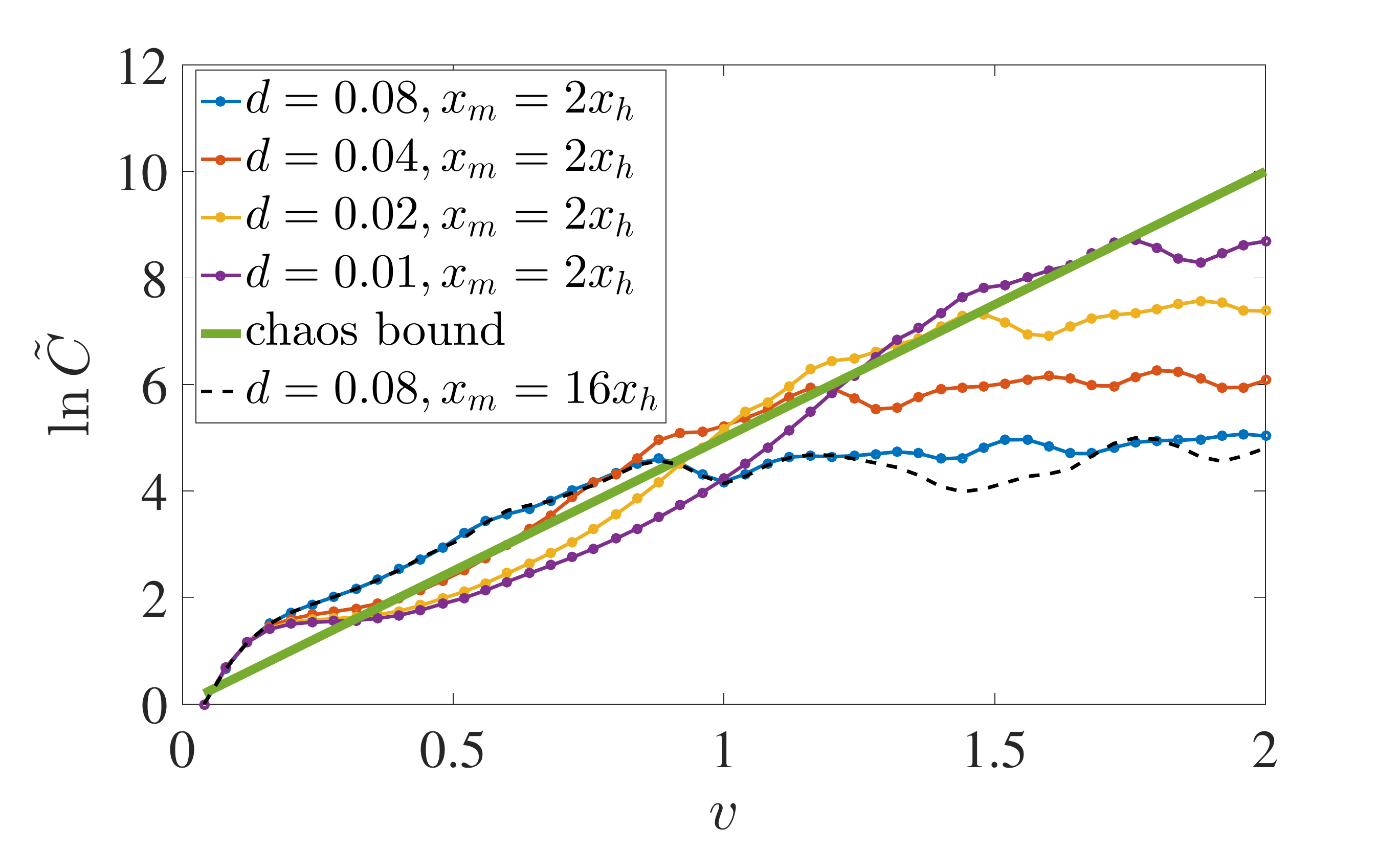}
\caption{The numerical simulations on evolution of OTOC when
$l_0=x_h/15,~~n_0=(x_h+2l_0)/d$. The green solid line
is fitted based on Eq.~\eqref{otocbd1}, where the Lyapunov exponent
$\lambda_L=2\pi T_H$. Here we choose $x_h=10$. We numerically simulate the results for time $v=0.02\sim2$ and $d=0.01,0.02,0.04$ and 0.08. $x_m$ is the position of AdS boundary. The numerical results are not sensitive to the value of cut-off boundary in the region of exponential growth. }\label{figotoc1}
\end{figure}

The time-evolution of $C(v)$ is obtained numerically. The results are shown in Fig.~\ref{figotoc1}.
For convenience, we define  $\tilde{C}(v)=C(v)/C(0.02)$, which does not change the slop of $\ln C(v)$.  For the numerical details, one can refer to the appendix.
We can observe that $C(v)$  exponentially growes approximately in
the early time. The slop
of the fitting curve is found to be $2\pi T_H$
approximately. The chaos bound~\eqref{otocbd2} is saturated approximately.

In the pure gravity theory, the effective  degree of freedom will be proportional to $G^{-2}$~\cite{Maldacena2016}, where $G$ is the Newton's constant. Here we neglect the backreaction of matters on geometry, which means the limit of $G\rightarrow0$. Thus, in principle, the OTOC will increase forever, i.e., $t_*\rightarrow\infty$. However, as we here use the lattice model, the operators and their commutators are bounded and so exponential growth will stop at a finite time. We study how the $C(v)$ depends on the discrete distance $d$. The results show that the time scale of exponential growth will increase if we decrease $d$ but fix the horizon radius $x_h$ and distribution width $l_0$ of $\hat{N}_{n_0}$. This suggests that the  time scale of exponential growth will become infinity in the continuous limit $d\rightarrow0$, as expected.

Strictly speaking, to claim a system to be chaotic, either classically or quantum mechanically, the positive Lyapunov exponent is necessary but not sufficient. The positive Lyapunov exponent only indicates the sensitivity  to the initial perturbations, which is the necessary condition of chaos. For example, in the classical case, we also require that the trajectory is dense in a neighborhood of phase space (i.e., ergodic). However, the linear analysis is enough to help us to find the Lyapunov exponents both in the classical case and quantum mechanical case. This can be understood by recalling the standard method in computing the Lyapunov exponent of classical chaotic systems. Thus, a linearized theory in a black hole background is enough to  check the ``chaos bound'' (it may be more suitable to call it ``bound on Lyapunov exponent'').

\begin{figure}
\centering
\includegraphics[width=0.3\textwidth]{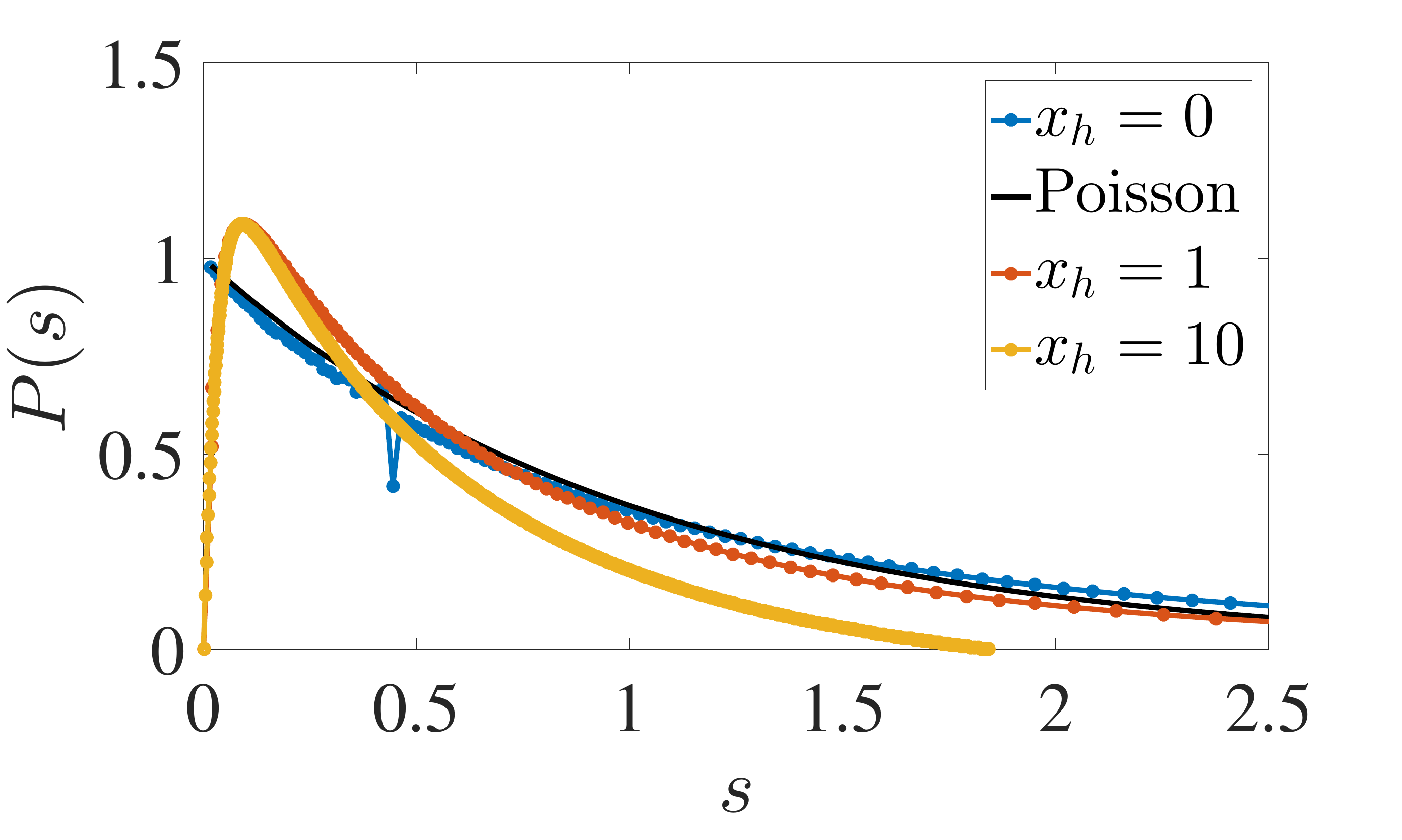}
\caption{Distribution of ``nearest-neighbor level spacings'' when the hopping $\kappa_n$ are given by a pure AdS spacetime (blue) and black holes (red and yellow), respectively. The black line is given by Poisson distribution $P(s)=e^{-s}$. We choose $d=0.02$ for all cases, $\{x_h=0,x_m=10\}$ for pure AdS case and $\{x_h=1,10, x_m=5x_h\}$ for the black hole cases. Here $x_m$ is the cut-off AdS boundary. An ``integrable-nonintegrale phase transition'' occurs when $x_n\neq0$. }\label{figotoc2}
\end{figure}

In order to check if the models~\eqref{deaHm2}, \eqref{deaHm3} and \eqref{deaHm4} really contain chaotic behaviors when the coupling constants are given by a black hole metric, we study the statistics of ``nearest-neighbor level spacing'', which is an other characteristic quantity of chaotic system. We denote the energy levels of outside Hamiltonian to be $E_i$ with $E_i<E_{i+1}$ , which are obtained by directly diagonalizing the Hamiltonian numerically (the cut-off in high energy levels is needed as high energy levels have low accuracy and are not trustworthy in physics). Assume that $\Delta$ to be the mean value of $E_{i+1}-E_i$ and $\mathcal{N}$ to be the total number of energy levels. Then we define $\mathcal{N}P(s)\delta s$ to be number of energy levels $E_i$ which satisfy $s\leq (E_{i+1}-E_i)/\Delta\leq s+\delta s$. The function $P(s)$ is called `nearest-neighbor level spacing'' function.
It has been shown that if the system is integrable, $P(s)$ satisfies Poisson statistics $P(s)=e^{-s}$~\cite{GUHR1998189}. If the system is chaotic, $P(s)$ will deviate from the Poisson statistics. For Gaussian orthogonal ensemble or Gaussian unitrary ensemble, $P(s)$ is given by Wigner distribution. For other general cases, the $P(s)$ may be given by general Brody distribution approximately~\cite{Jafarizadeh:2012da}. In  Fig~\eqref{figotoc2} we show the numerical results about $P(s)$. For the case that $x_h=0$, the effective spacetime has no black hole and we find that $P(s)$ is given by a Poisson statistics approximately. However, once $x_h\neq0$, we find that a dramatic change happens and $P(s)$ is no longer a Poisson distribution, which suggests that the system is not integrable.

From the viewpoint of quantum many-body theory, Hamiltonians shown in Eqs.~\eqref{deaHm}, \eqref{deaHm2} and \eqref{deaHm4} contain only nearest hopping and quadratic interactions no matter how to design coupling constants. When we design  the coupling constant by setting $x_h\neq0$, it is not easy to understand why these models can exhibit exponential growth of OTOC and why the systems have ``integrable-nonintegrale phase transition''. However, if we use our framework to convert them into an effective black hole, we immediately understand these properties. When $x_h\neq0$, a black hole metric is encoded into the sit-dependent coupling $\kappa_{n}$ and so the system has exponential growth of OTOC and signal of chaos.
This offers an example about how to use our models to get some insights  about the quantum many-body systems from the back hole physics.

\section{Summary}
In summary, we have  shown that a massless scalar/Dirac field in the static 1+1 dimensional curved spacetime can be simulated by some basic models in condensed matter physics: the bosonic hopping model, free Hubbard model and XY model. We suggested a possible experimental realization in trapped ions system for this analogue gravity model. As some applications of the analogue gravity model, we have numerically  shown that this model can be used to simulate Hawking radiation. We have also checked the quantum chaos behave of black hole and verified that  black hole is one of the most chaotic systems and has the fastest scrambling in nature.  
These are predictions of AdS/CFT correspondence. In this sense our model provides the possibility to test experimentally the correspondence. In addition,
our results show the site-dependent hopping is one-to-one related to  spacetime point  of curved background. By a concrete example, we showed how this framework can help us  get some insights about the quantum many-body systems from the back hole physics. This not only provides new motivation to study the site-dependent
hopping model, but also indicates a large number of applications in the analogue gravity model. These will bring us new viewpoints and  phenomena of quantum many-body systems, and also will enlighten us to deeply understand the nature of gravity.

\appendix
\section{Tunneling rate and Hawking temperature}\label{app1}
In this appendix, let us show how to use the picture ``quantum tunneling'' to obtain the tunneling rate and the Hawking temperature of black hole.

To obtain the tunneling rate, we need to find the solution Eq.~\eqref{indepnulleq2} with the energy $E$. By a variable transformation
\begin{equation}\label{u2tildeu}
  w={\phi}/\sqrt{f}
\end{equation}
we can find that
\begin{equation}\label{indepnulleq3}
  \partial_v{\phi}=-\frac{f}2\partial_x{\phi}\,.
\end{equation}
The positive energy (measured by $v$) solution is
\begin{equation}\label{solutionu1}
  {\phi}=\phi_0\exp\left[-iE\left(v-2\int\frac{\td x}{f(x)}\right)\right]\,.
\end{equation}
As $f(x)=0$ at the horizon, this solution is not continuous at the horizon. Let us separate the integration in the above equation as follows
\begin{equation}\label{intelog1}
\begin{split}
  \int\frac{\td x}{f(x)}&=\int\left(\frac{1}{f(x)}-\frac{1}{2g_h(x-x_h)}\right)\td x+\frac1{2g_h}\ln|x-x_h|\\
  &=F(x)+\frac1{2g_h}\ln|x-x_h|\,.
  \end{split}
\end{equation}
The function $F(x)$ is continuous at the horizon. The divergence has been absorbed into the logarithm function.
The solution~\eqref{solutionu1} can be separated into two pieces,
%
\begin{equation}\label{solutionu2a}
  {\phi}={\phi}_1\exp\left\{-i E \left[v-2F(x)-\frac1{g_h}\ln(x_h-x)\right]\right\}
\end{equation}
for $x<x_h$ and
\begin{equation}\label{solutionu2b}
  {\phi}={\phi}_2\exp\left\{-i E \left[v-2F(x)-\frac1{g_h}\ln(x-x_h)\right]\right\}
\end{equation}
for $x>x_h$.
The tunneling rate then reads,
\begin{equation}\label{defGamma}
  \Gamma:=\frac{|{\phi}_2|^2}{|{\phi}_1|^2}\,.
\end{equation}
Following the argument in Ref.~\cite{Damour:1976jd}, the two pieces of the solution in Eqs.~\eqref{solutionu2a} and \eqref{solutionu2b} should be connected continuously under the bottom half of complex plane. Treat the piece of $x<x_h$ as the starting point and analytically continue it into the region of $x>x_h$, the logarithm function in Eq.~\eqref{solutionu2a} will obtain an additional phase factor and so we can obtain the following relationship
$${\phi}_1\exp\left(-\frac{\pi E}{g_h}\right)={\phi}_2\,.$$
Take it into Eq.~\eqref{defGamma} and we then obtain
\begin{equation}\label{thermal0}
  \Gamma=\exp\left(-\frac{2\pi E}{g_h}\right)=\exp\left(-\frac{E}{T_H}\right).
\end{equation}
As expected, the tunneling rate and energy satisfy the blackbody spectrum and the temperature is just given by $T_H=g_h/(2\pi)$.

In physics, Eq.~\eqref{solutionu1} implies an infinite momentum at horizon and so will break down our condition for discretization. This  belongs to the question named ``trans-Planckian problem'', which widely exists in all discussions of Hawking radiation. A particle emitted from a black hole with a finite frequency (measured at infinity), if traced back to the horizon, must have an infinite momentum, and therefore a trans-Planckian wavelength. The trans-Planckian problem is a mathematical artifact of horizon calculations. In all analogue models, when the emitted particle is near the ``horizon'', the smooth approximation is invalid and so a truncation is needed. However, it has been showed that the details of truncation will not change the behavior of Hawking radiation in ``low energy'' region (the ``low energy'' means the energy is low at infinity), see Ref.~\cite{Unruh:1994je}, for example.


\section{Details of numerical simulations on Hawking radiation}
Let us first explain how to make the numerical simulation on Hawking radiation. We take $f(x)=\alpha\tanh x$
Then we can see that the hopping amplitude reads
$$\kappa_n=\frac{\alpha\tanh[(n-1/2)d]}{4d}\,.$$
There is a horizon at $x=x_h=0$ with the Hawking temperature $T_H=\alpha/(4\pi)$. It is worth noting that $\kappa_n\neq0$ at horizon though $f(x)$ is zero at horizon.
The numerical computation needs a finite cut-off $n=-L,-L+1,\cdots,L-1,L$. To match this cut-off, we have to set hopping amplitude $\kappa_n$ such that
$$\kappa_{n}=0,~\text{if}~n\geq L~\text{or}~n\leq-L\,.$$
Without lose of generality, we set $\mu=0$ as the total particle number is conserved.


The Hamiltonians for inner region and outer region are
\begin{equation}\label{defHIONIO}
  \Hm_{\text{in}}=-\sum_{n=-L}^{-1}\kappa_n\left(\ha_n^\dagger\ha_{n-1}+\ha_{n-1}^\dagger\ha_{n}\right), ~~
\end{equation}
and
\begin{equation}\label{defHIONIO2}
  \Hm_{\text{out}}=-\sum_{n=2}^{L}\kappa_n\left(\ha_n^\dagger\ha_{n-1}+\ha_{n-1}^\dagger\ha_{n}\right).
\end{equation}
Note that the total Hamiltonian is not the sum of inner part and outer part. In fact, we have
\begin{equation}\label{totHmHiHo}
  \Hm=\Hm_{\text{in}}+\Hm_{\text{out}}+\Hm_{0}\,.
\end{equation}
Here $\Hm_{0}$ is the contribution at the horizon
\begin{equation}\label{horionHm}
  \Hm_{0}=-\kappa_0(\ha_0^\dagger\ha_{-1}+\ha_{-1}^\dagger\ha_{0})-\kappa_1(\ha_1^\dagger\ha_{0}+\ha_{0}^\dagger\ha_{1})\,,
\end{equation}
which mixes the inner region and outer region.

Assume $N$ to be the total particle number. It is difficult to simulate the dynamics for large $N$ and $L$. For example,  in the case $2L+1=N=13$, the dimension of total Hilbert space is  $D\approx5\times10^{6}$.
To simplify the issue in numerical algorithm, let us choose $N=1$ and so the dimension of Hilbert space is $D=2L+1$.
In this case, we can choose the eigenvectors of $\ha_n^\dagger\ha_n$ as the basic vectors of Hilbert space
\begin{equation}\label{basicvector}
  \begin{split}
  &|e_{-L}\rangle=(1,0,\cdots,0)^T,\\
  &|e_{-L+1}\rangle=(0,1,0,\cdots,0)^T,\\
  &\cdots,\\
  &|e_L\rangle=(0,0,0,\cdots,1)^T\,,
  \end{split}
\end{equation}
which satisfy
\begin{equation}\label{aaelement1}
  \langle e_l|\ha_n^\dagger\ha_{n}|e_k\rangle=\delta_{nl}\delta_{nk}
\end{equation}
and
\begin{equation}\label{aaelement2}
  \langle e_l|\ha_n^\dagger\ha_{n-1}|e_k\rangle=\delta_{n-1,k}\delta_{l,n}\,.
\end{equation}
Then we can write down the matrix elements of Hamiltonian.
\begin{equation}\label{matrixHm1}
(\Hm_{\text{in}})_{l,k}=\left\{
\begin{split}
  &-\kappa_l(\delta_{k,l-1}+\delta_{l,k-1}),~~~k,l\leq-1\\
  &0,~~k,l>-1
  \end{split}
  \right.
\end{equation}
\begin{equation}\label{matrixHm2}
(\Hm_{\text{out}})_{l,k}=\left\{
\begin{split}
  &-\kappa_l(\delta_{k,l-1}+\delta_{l,k-1}),~~~k,l\geq2\\
  &0,~~k,l<2
  \end{split}
  \right.
\end{equation}
and
\begin{equation}\label{matrixHm3}
(\Hm)_{l,k}=-\kappa_l(\delta_{k,l-1}+\delta_{l,k-1})
\end{equation}
For a given initial state $|\Psi(0)\rangle$, the time evolutional state is given by $|\Psi(v)\rangle=e^{-i\Hm v}|\Psi(0)\rangle$. In  Fig. 1,  we take parameters $d=0.1,L=300$ and $\alpha=10$.
The results are similar if we choose $N=2$. Due to the technical difficulties addressed above, we cannot explore larger $N$. Roughly speaking, as we consider the free theory in fixed background, many particles can be understood as a collection of single particle. Thus, the simplification here does not loss the essential physics.

\section{Details of numerical simulations on OTOC}
The simulation on OTOC is similar. We take $f(x)=x^2(1-x_h/x)$. Then we can see that the hopping reads
$$\kappa_n=\frac{f[(n-1/2)d]}{4d}\,.$$
There is a horizon at $x=x_h$ with the Hawking temperature $T_H=x_h/(4\pi)$. In this case we make the cut-off in the following way
$$n=1,2,\cdots,2L+1,~~~\text{with}~Ld=x_h\,.$$
%

Similar to the case in Hawking radiation, if we choose $N=1$, we still have Eqs.~\eqref{aaelement1} and \eqref{aaelement2}. Then we can write down the matrix elements of Hamiltonian
\begin{equation}\label{matrixHm3c}
(\Hm)_{l,k}=-\kappa_l(\delta_{k,l-1}+\delta_{l,k-1})
\end{equation}
and
\begin{equation}\label{matrixN1}
  (\ha_n^\dagger\ha_n)_{kl}=\delta_{nl}\delta_{nk}\,.
\end{equation}
For a given initial state $|\Psi(0)\rangle$, the time evolutional state is given by $|\Psi(v)\rangle=e^{-i\Hm v}|\Psi(0)\rangle$ and $\hat{N}_n(v)=\exp(-i\Hm v)\hat{N}_n\exp(i\Hm v)$. Then we can obtain the OTOC in Eq. (27).

\begin{acknowledgments}
R.-G. Cai was supported in part by the National
Natural Science Foundation of China Grants Nos.
11690022, 11821505, 11851302, 11947302, 11745006,  and by the Strategic Priority Research
Program of CAS Grant No.XDB23030100, and by the Key Research Program
of Frontier Sciences of CAS. H. Liu was supported by the National
Natural Science Foundation of China Grants No.11690033. L. Luo is a
member of the Indiana University Center for Spacetime Symmetries
(IUCSS). L. Luo received supports from National Natural Science
Foundation of China under Grant No. 11774436, Sun Yat-sen University
Discipline Construction Fund, Sun Yat-sen University Three Major
Construction Fund, and Guangdong Province Pearl River Youth Talent Fund.
\end{acknowledgments}

\bibliography{Hubbard1}

\begin{thebibliography}{47}%
\makeatletter
\providecommand \@ifxundefined [1]{%
 \@ifx{#1\undefined}
}%
\providecommand \@ifnum [1]{%
 \ifnum #1\expandafter \@firstoftwo
 \else \expandafter \@secondoftwo
 \fi
}%
\providecommand \@ifx [1]{%
 \ifx #1\expandafter \@firstoftwo
 \else \expandafter \@secondoftwo
 \fi
}%
\providecommand \natexlab [1]{#1}%
\providecommand \enquote  [1]{``#1''}%
\providecommand \bibnamefont  [1]{#1}%
\providecommand \bibfnamefont [1]{#1}%
\providecommand \citenamefont [1]{#1}%
\providecommand \href@noop [0]{\@secondoftwo}%
\providecommand \href [0]{\begingroup \@sanitize@url \@href}%
\providecommand \@href[1]{\@@startlink{#1}\@@href}%
\providecommand \@@href[1]{\endgroup#1\@@endlink}%
\providecommand \@sanitize@url [0]{\catcode `\\12\catcode `\$12\catcode
  `\&12\catcode `\#12\catcode `\^12\catcode `\_12\catcode `\%12\relax}%
\providecommand \@@startlink[1]{}%
\providecommand \@@endlink[0]{}%
\providecommand \url  [0]{\begingroup\@sanitize@url \@url }%
\providecommand \@url [1]{\endgroup\@href {#1}{\urlprefix }}%
\providecommand \urlprefix  [0]{URL }%
\providecommand \Eprint [0]{\href }%
\providecommand \doibase [0]{http://dx.doi.org/}%
\providecommand \selectlanguage [0]{\@gobble}%
\providecommand \bibinfo  [0]{\@secondoftwo}%
\providecommand \bibfield  [0]{\@secondoftwo}%
\providecommand \translation [1]{[#1]}%
\providecommand \BibitemOpen [0]{}%
\providecommand \bibitemStop [0]{}%
\providecommand \bibitemNoStop [0]{.\EOS\space}%
\providecommand \EOS [0]{\spacefactor3000\relax}%
\providecommand \BibitemShut  [1]{\csname bibitem#1\endcsname}%
\let\auto@bib@innerbib\@empty
\bibitem [{\citenamefont {Jacobson}(2003)}]{Jacobson:2003vx}%
  \BibitemOpen
  \bibfield  {author} {\bibinfo {author} {\bibfnamefont {Ted}\ \bibnamefont
  {Jacobson}},\ }\bibfield  {title} {\enquote {\bibinfo {title} {{Introduction
  to quantum fields in curved space-time and the Hawking effect}},}\ }in\ \href
  {\doibase 10.1007/0-387-24992-3_2} {\emph {\bibinfo {booktitle} {{Lectures on
  quantum gravity. Proceedings, School of Quantum Gravity, Valdivia, Chile,
  January 4-14, 2002}}}}\ (\bibinfo {year} {2003})\ pp.\ \bibinfo {pages}
  {39--89},\ \Eprint {http://arxiv.org/abs/gr-qc/0308048} {arXiv:gr-qc/0308048
  [gr-qc]} \BibitemShut {NoStop}%
\bibitem [{\citenamefont {Hollands}\ and\ \citenamefont
  {Wald}(2015)}]{Hollands:2014eia}%
  \BibitemOpen
  \bibfield  {author} {\bibinfo {author} {\bibfnamefont {Stefan}\ \bibnamefont
  {Hollands}}\ and\ \bibinfo {author} {\bibfnamefont {Robert~M.}\ \bibnamefont
  {Wald}},\ }\bibfield  {title} {\enquote {\bibinfo {title} {{Quantum fields in
  curved spacetime}},}\ }\href {\doibase 10.1016/j.physrep.2015.02.001}
  {\bibfield  {journal} {\bibinfo  {journal} {Phys. Rept.}\ }\textbf {\bibinfo
  {volume} {574}},\ \bibinfo {pages} {1--35} (\bibinfo {year} {2015})},\
  \Eprint {http://arxiv.org/abs/1401.2026} {arXiv:1401.2026 [gr-qc]}
  \BibitemShut {NoStop}%
\bibitem [{\citenamefont {Unruh}(1981)}]{PhysRevLett.46.1351}%
  \BibitemOpen
  \bibfield  {author} {\bibinfo {author} {\bibfnamefont {W.~G.}\ \bibnamefont
  {Unruh}},\ }\bibfield  {title} {\enquote {\bibinfo {title} {Experimental
  black-hole evaporation?}}\ }\href {\doibase 10.1103/PhysRevLett.46.1351}
  {\bibfield  {journal} {\bibinfo  {journal} {Phys. Rev. Lett.}\ }\textbf
  {\bibinfo {volume} {46}},\ \bibinfo {pages} {1351--1353} (\bibinfo {year}
  {1981})}\BibitemShut {NoStop}%
\bibitem [{\citenamefont {Unruh}(1995)}]{Unruh:1994je}%
  \BibitemOpen
  \bibfield  {author} {\bibinfo {author} {\bibfnamefont {W.~G.}\ \bibnamefont
  {Unruh}},\ }\bibfield  {title} {\enquote {\bibinfo {title} {{Sonic analog of
  black holes and the effects of high frequencies on black hole
  evaporation}},}\ }\href {\doibase 10.1103/PhysRevD.51.2827} {\bibfield
  {journal} {\bibinfo  {journal} {Phys. Rev.}\ }\textbf {\bibinfo {volume}
  {D51}},\ \bibinfo {pages} {2827--2838} (\bibinfo {year} {1995})}\BibitemShut
  {NoStop}%
\bibitem [{\citenamefont {Weinfurtner}\ \emph {et~al.}(2011)\citenamefont
  {Weinfurtner}, \citenamefont {Tedford}, \citenamefont {Penrice},
  \citenamefont {Unruh},\ and\ \citenamefont
  {Lawrence}}]{PhysRevLett.106.021302}%
  \BibitemOpen
  \bibfield  {author} {\bibinfo {author} {\bibfnamefont {Silke}\ \bibnamefont
  {Weinfurtner}}, \bibinfo {author} {\bibfnamefont {Edmund~W.}\ \bibnamefont
  {Tedford}}, \bibinfo {author} {\bibfnamefont {Matthew C.~J.}\ \bibnamefont
  {Penrice}}, \bibinfo {author} {\bibfnamefont {William~G.}\ \bibnamefont
  {Unruh}}, \ and\ \bibinfo {author} {\bibfnamefont {Gregory~A.}\ \bibnamefont
  {Lawrence}},\ }\bibfield  {title} {\enquote {\bibinfo {title} {Measurement of
  stimulated hawking emission in an analogue system},}\ }\href {\doibase
  10.1103/PhysRevLett.106.021302} {\bibfield  {journal} {\bibinfo  {journal}
  {Phys. Rev. Lett.}\ }\textbf {\bibinfo {volume} {106}},\ \bibinfo {pages}
  {021302} (\bibinfo {year} {2011})}\BibitemShut {NoStop}%
\bibitem [{\citenamefont {Garay}\ \emph {et~al.}(2000)\citenamefont {Garay},
  \citenamefont {Anglin}, \citenamefont {Cirac},\ and\ \citenamefont
  {Zoller}}]{PhysRevLett.85.4643}%
  \BibitemOpen
  \bibfield  {author} {\bibinfo {author} {\bibfnamefont {L.~J.}\ \bibnamefont
  {Garay}}, \bibinfo {author} {\bibfnamefont {J.~R.}\ \bibnamefont {Anglin}},
  \bibinfo {author} {\bibfnamefont {J.~I.}\ \bibnamefont {Cirac}}, \ and\
  \bibinfo {author} {\bibfnamefont {P.}~\bibnamefont {Zoller}},\ }\bibfield
  {title} {\enquote {\bibinfo {title} {Sonic analog of gravitational black
  holes in bose-einstein condensates},}\ }\href {\doibase
  10.1103/PhysRevLett.85.4643} {\bibfield  {journal} {\bibinfo  {journal}
  {Phys. Rev. Lett.}\ }\textbf {\bibinfo {volume} {85}},\ \bibinfo {pages}
  {4643--4647} (\bibinfo {year} {2000})}\BibitemShut {NoStop}%
\bibitem [{\citenamefont {Steinhauer}(2016)}]{Steinhauer2016}%
  \BibitemOpen
  \bibfield  {author} {\bibinfo {author} {\bibfnamefont {Jeff}\ \bibnamefont
  {Steinhauer}},\ }\bibfield  {title} {\enquote {\bibinfo {title} {Observation
  of quantum hawking radiation and its entanglement in an analogue black
  hole},}\ }\href {\doibase 10.1038/nphys3863} {\bibfield  {journal} {\bibinfo
  {journal} {Nature Physics}\ }\textbf {\bibinfo {volume} {12}},\ \bibinfo
  {pages} {959--965} (\bibinfo {year} {2016})}\BibitemShut {NoStop}%
\bibitem [{\citenamefont {de~Nova}\ \emph {et~al.}(2019)\citenamefont
  {de~Nova}, \citenamefont {Golubkov}, \citenamefont {Kolobov},\ and\
  \citenamefont {Steinhauer}}]{MuozdeNova2019}%
  \BibitemOpen
  \bibfield  {author} {\bibinfo {author} {\bibfnamefont {Juan
  Ram{\'{o}}n~Mu{\~{n}}oz}\ \bibnamefont {de~Nova}}, \bibinfo {author}
  {\bibfnamefont {Katrine}\ \bibnamefont {Golubkov}}, \bibinfo {author}
  {\bibfnamefont {Victor~I.}\ \bibnamefont {Kolobov}}, \ and\ \bibinfo {author}
  {\bibfnamefont {Jeff}\ \bibnamefont {Steinhauer}},\ }\bibfield  {title}
  {\enquote {\bibinfo {title} {Observation of thermal hawking radiation and its
  temperature in an analogue black hole},}\ }\href {\doibase
  10.1038/s41586-019-1241-0} {\bibfield  {journal} {\bibinfo  {journal}
  {Nature}\ }\textbf {\bibinfo {volume} {569}},\ \bibinfo {pages} {688--691}
  (\bibinfo {year} {2019})}\BibitemShut {NoStop}%
\bibitem [{\citenamefont {Sheng}\ \emph {et~al.}(2013)\citenamefont {Sheng},
  \citenamefont {Liu}, \citenamefont {Wang}, \citenamefont {Zhu},\ and\
  \citenamefont {Genov}}]{Sheng2013}%
  \BibitemOpen
  \bibfield  {author} {\bibinfo {author} {\bibfnamefont {C.}~\bibnamefont
  {Sheng}}, \bibinfo {author} {\bibfnamefont {H.}~\bibnamefont {Liu}}, \bibinfo
  {author} {\bibfnamefont {Y.}~\bibnamefont {Wang}}, \bibinfo {author}
  {\bibfnamefont {S.~N.}\ \bibnamefont {Zhu}}, \ and\ \bibinfo {author}
  {\bibfnamefont {D.~A.}\ \bibnamefont {Genov}},\ }\bibfield  {title} {\enquote
  {\bibinfo {title} {Trapping light by mimicking gravitational lensing},}\
  }\href {\doibase 10.1038/nphoton.2013.247} {\bibfield  {journal} {\bibinfo
  {journal} {Nature Photonics}\ }\textbf {\bibinfo {volume} {7}},\ \bibinfo
  {pages} {902--906} (\bibinfo {year} {2013})}\BibitemShut {NoStop}%
\bibitem [{\citenamefont {Bekenstein}\ \emph {et~al.}(2017)\citenamefont
  {Bekenstein}, \citenamefont {Kabessa}, \citenamefont {Sharabi}, \citenamefont
  {Tal}, \citenamefont {Engheta}, \citenamefont {Eisenstein}, \citenamefont
  {Agranat},\ and\ \citenamefont {Segev}}]{Bekenstein2017}%
  \BibitemOpen
  \bibfield  {author} {\bibinfo {author} {\bibfnamefont {Rivka}\ \bibnamefont
  {Bekenstein}}, \bibinfo {author} {\bibfnamefont {Yossef}\ \bibnamefont
  {Kabessa}}, \bibinfo {author} {\bibfnamefont {Yonatan}\ \bibnamefont
  {Sharabi}}, \bibinfo {author} {\bibfnamefont {Or}~\bibnamefont {Tal}},
  \bibinfo {author} {\bibfnamefont {Nader}\ \bibnamefont {Engheta}}, \bibinfo
  {author} {\bibfnamefont {Gadi}\ \bibnamefont {Eisenstein}}, \bibinfo {author}
  {\bibfnamefont {Aharon~J.}\ \bibnamefont {Agranat}}, \ and\ \bibinfo {author}
  {\bibfnamefont {Mordechai}\ \bibnamefont {Segev}},\ }\bibfield  {title}
  {\enquote {\bibinfo {title} {Control of light by curved space in nanophotonic
  structures},}\ }\href {\doibase 10.1038/s41566-017-0008-0} {\bibfield
  {journal} {\bibinfo  {journal} {Nature Photonics}\ }\textbf {\bibinfo
  {volume} {11}},\ \bibinfo {pages} {664--670} (\bibinfo {year}
  {2017})}\BibitemShut {NoStop}%
\bibitem [{\citenamefont {Drori}\ \emph {et~al.}(2019)\citenamefont {Drori},
  \citenamefont {Rosenberg}, \citenamefont {Bermudez}, \citenamefont
  {Silberberg},\ and\ \citenamefont {Leonhardt}}]{PhysRevLett.122.010404}%
  \BibitemOpen
  \bibfield  {author} {\bibinfo {author} {\bibfnamefont {Jonathan}\
  \bibnamefont {Drori}}, \bibinfo {author} {\bibfnamefont {Yuval}\ \bibnamefont
  {Rosenberg}}, \bibinfo {author} {\bibfnamefont {David}\ \bibnamefont
  {Bermudez}}, \bibinfo {author} {\bibfnamefont {Yaron}\ \bibnamefont
  {Silberberg}}, \ and\ \bibinfo {author} {\bibfnamefont {Ulf}\ \bibnamefont
  {Leonhardt}},\ }\bibfield  {title} {\enquote {\bibinfo {title} {Observation
  of stimulated hawking radiation in an optical analogue},}\ }\href {\doibase
  10.1103/PhysRevLett.122.010404} {\bibfield  {journal} {\bibinfo  {journal}
  {Phys. Rev. Lett.}\ }\textbf {\bibinfo {volume} {122}},\ \bibinfo {pages}
  {010404} (\bibinfo {year} {2019})}\BibitemShut {NoStop}%
\bibitem [{\citenamefont {Sheng}\ \emph {et~al.}(2018)\citenamefont {Sheng},
  \citenamefont {Liu}, \citenamefont {Chen},\ and\ \citenamefont
  {Zhu}}]{Sheng2018}%
  \BibitemOpen
  \bibfield  {author} {\bibinfo {author} {\bibfnamefont {Chong}\ \bibnamefont
  {Sheng}}, \bibinfo {author} {\bibfnamefont {Hui}\ \bibnamefont {Liu}},
  \bibinfo {author} {\bibfnamefont {Huanyang}\ \bibnamefont {Chen}}, \ and\
  \bibinfo {author} {\bibfnamefont {Shining}\ \bibnamefont {Zhu}},\ }\bibfield
  {title} {\enquote {\bibinfo {title} {Definite photon deflections
  of~topological~defects in metasurfaces and symmetry-breaking phase
  transitions with material loss},}\ }\href {\doibase
  10.1038/s41467-018-06718-9} {\bibfield  {journal} {\bibinfo  {journal}
  {Nature Communications}\ }\textbf {\bibinfo {volume} {9}} (\bibinfo {year}
  {2018}),\ 10.1038/s41467-018-06718-9}\BibitemShut {NoStop}%
\bibitem [{\citenamefont {Zhong}\ \emph {et~al.}(2018)\citenamefont {Zhong},
  \citenamefont {Li}, \citenamefont {Liu},\ and\ \citenamefont
  {Zhu}}]{PhysRevLett.120.243901}%
  \BibitemOpen
  \bibfield  {author} {\bibinfo {author} {\bibfnamefont {Fan}\ \bibnamefont
  {Zhong}}, \bibinfo {author} {\bibfnamefont {Jensen}\ \bibnamefont {Li}},
  \bibinfo {author} {\bibfnamefont {Hui}\ \bibnamefont {Liu}}, \ and\ \bibinfo
  {author} {\bibfnamefont {Shining}\ \bibnamefont {Zhu}},\ }\bibfield  {title}
  {\enquote {\bibinfo {title} {Controlling surface plasmons through covariant
  transformation of the spin-dependent geometric phase between curved
  metamaterials},}\ }\href {\doibase 10.1103/PhysRevLett.120.243901} {\bibfield
   {journal} {\bibinfo  {journal} {Phys. Rev. Lett.}\ }\textbf {\bibinfo
  {volume} {120}},\ \bibinfo {pages} {243901} (\bibinfo {year}
  {2018})}\BibitemShut {NoStop}%
\bibitem [{\citenamefont {Pedernales}\ \emph {et~al.}(2018)\citenamefont
  {Pedernales}, \citenamefont {Beau}, \citenamefont {Pittman}, \citenamefont
  {Egusquiza}, \citenamefont {Lamata}, \citenamefont {Solano},\ and\
  \citenamefont {del Campo}}]{Pedernales:2017eue}%
  \BibitemOpen
  \bibfield  {author} {\bibinfo {author} {\bibfnamefont {J.~S.}\ \bibnamefont
  {Pedernales}}, \bibinfo {author} {\bibfnamefont {M.}~\bibnamefont {Beau}},
  \bibinfo {author} {\bibfnamefont {S.~M.}\ \bibnamefont {Pittman}}, \bibinfo
  {author} {\bibfnamefont {I.~L.}\ \bibnamefont {Egusquiza}}, \bibinfo {author}
  {\bibfnamefont {L.}~\bibnamefont {Lamata}}, \bibinfo {author} {\bibfnamefont
  {E.}~\bibnamefont {Solano}}, \ and\ \bibinfo {author} {\bibfnamefont
  {A.}~\bibnamefont {del Campo}},\ }\bibfield  {title} {\enquote {\bibinfo
  {title} {{Dirac Equation in ( 1+1 )-Dimensional Curved Spacetime and the
  Multiphoton Quantum Rabi Model}},}\ }\href {\doibase
  10.1103/PhysRevLett.120.160403} {\bibfield  {journal} {\bibinfo  {journal}
  {Phys. Rev. Lett.}\ }\textbf {\bibinfo {volume} {120}},\ \bibinfo {pages}
  {160403} (\bibinfo {year} {2018})},\ \Eprint
  {http://arxiv.org/abs/1707.07520} {arXiv:1707.07520 [quant-ph]} \BibitemShut
  {NoStop}%
\bibitem [{\citenamefont {Barcelo}\ \emph {et~al.}(2005)\citenamefont
  {Barcelo}, \citenamefont {Liberati},\ and\ \citenamefont
  {Visser}}]{Barcelo:2005fc}%
  \BibitemOpen
  \bibfield  {author} {\bibinfo {author} {\bibfnamefont {Carlos}\ \bibnamefont
  {Barcelo}}, \bibinfo {author} {\bibfnamefont {Stefano}\ \bibnamefont
  {Liberati}}, \ and\ \bibinfo {author} {\bibfnamefont {Matt}\ \bibnamefont
  {Visser}},\ }\bibfield  {title} {\enquote {\bibinfo {title} {{Analogue
  gravity}},}\ }\href {\doibase 10.12942/lrr-2005-12} {\bibfield  {journal}
  {\bibinfo  {journal} {Living Rev. Rel.}\ }\textbf {\bibinfo {volume} {8}},\
  \bibinfo {pages} {12} (\bibinfo {year} {2005})},\ \bibinfo {note} {[Living
  Rev. Rel.14,3(2011)]},\ \Eprint {http://arxiv.org/abs/gr-qc/0505065}
  {arXiv:gr-qc/0505065 [gr-qc]} \BibitemShut {NoStop}%
\bibitem [{\citenamefont {Faccio}\ \emph {et~al.}(2013)\citenamefont {Faccio},
  \citenamefont {Belgiorno}, \citenamefont {Cacciatori}, \citenamefont
  {Gorini}, \citenamefont {Liberati},\ and\ \citenamefont
  {Moschella}}]{9783319002668}%
  \BibitemOpen
  \bibfield  {author} {\bibinfo {author} {\bibfnamefont {Daniele}\ \bibnamefont
  {Faccio}}, \bibinfo {author} {\bibfnamefont {Francesco}\ \bibnamefont
  {Belgiorno}}, \bibinfo {author} {\bibfnamefont {Sergio}\ \bibnamefont
  {Cacciatori}}, \bibinfo {author} {\bibfnamefont {Vittorio}\ \bibnamefont
  {Gorini}}, \bibinfo {author} {\bibfnamefont {Stefano}\ \bibnamefont
  {Liberati}}, \ and\ \bibinfo {author} {\bibfnamefont {Ugo}\ \bibnamefont
  {Moschella}},\ }\href@noop {} {\emph {\bibinfo {title} {Analogue Gravity
  Phenomenology}}}\ (\bibinfo {year} {2013})\BibitemShut {NoStop}%
\bibitem [{\citenamefont {Barcel{\'{o}}}(2018)}]{Barcel2018}%
  \BibitemOpen
  \bibfield  {author} {\bibinfo {author} {\bibfnamefont {Carlos}\ \bibnamefont
  {Barcel{\'{o}}}},\ }\bibfield  {title} {\enquote {\bibinfo {title} {Analogue
  black-hole horizons},}\ }\href {\doibase 10.1038/s41567-018-0367-6}
  {\bibfield  {journal} {\bibinfo  {journal} {Nature Physics}\ }\textbf
  {\bibinfo {volume} {15}},\ \bibinfo {pages} {210--213} (\bibinfo {year}
  {2018})}\BibitemShut {NoStop}%
\bibitem [{\citenamefont {Tasaki}(1998)}]{Tasaki-1998}%
  \BibitemOpen
  \bibfield  {author} {\bibinfo {author} {\bibfnamefont {Hal}\ \bibnamefont
  {Tasaki}},\ }\bibfield  {title} {\enquote {\bibinfo {title} {The hubbard
  model - an introduction and selected rigorous results},}\ }\href {\doibase
  10.1088/0953-8984/10/20/004} {\bibfield  {journal} {\bibinfo  {journal}
  {Journal of Physics: Condensed Matter}\ }\textbf {\bibinfo {volume} {10}},\
  \bibinfo {pages} {4353--4378} (\bibinfo {year} {1998})}\BibitemShut {NoStop}%
\bibitem [{\citenamefont {Hu}\ \emph {et~al.}(2019)\citenamefont {Hu},
  \citenamefont {Feng}, \citenamefont {Zhang},\ and\ \citenamefont
  {Chin}}]{Hu:2018psq}%
  \BibitemOpen
  \bibfield  {author} {\bibinfo {author} {\bibfnamefont {Jiazhong}\
  \bibnamefont {Hu}}, \bibinfo {author} {\bibfnamefont {Lei}\ \bibnamefont
  {Feng}}, \bibinfo {author} {\bibfnamefont {Zhendong}\ \bibnamefont {Zhang}},
  \ and\ \bibinfo {author} {\bibfnamefont {Cheng}\ \bibnamefont {Chin}},\
  }\bibfield  {title} {\enquote {\bibinfo {title} {{Quantum simulation of Unruh
  radiation}},}\ }\href {\doibase 10.1038/s41567-019-0537-1} {\bibfield
  {journal} {\bibinfo  {journal} {Nature Phys.}\ }\textbf {\bibinfo {volume}
  {15}},\ \bibinfo {pages} {785--789} (\bibinfo {year} {2019})},\ \Eprint
  {http://arxiv.org/abs/1807.07504} {arXiv:1807.07504 [physics.atom-ph]}
  \BibitemShut {NoStop}%
\bibitem [{\citenamefont {Trautmann}\ and\ \citenamefont
  {Hauke}(2016)}]{Trautmann:2016cpg}%
  \BibitemOpen
  \bibfield  {author} {\bibinfo {author} {\bibfnamefont {N.}~\bibnamefont
  {Trautmann}}\ and\ \bibinfo {author} {\bibfnamefont {P.}~\bibnamefont
  {Hauke}},\ }\bibfield  {title} {\enquote {\bibinfo {title} {{Quantum
  simulation of the dynamical Casimir effect with trapped ions}},}\ }\href
  {\doibase 10.1088/1367-2630/18/4/043029} {\bibfield  {journal} {\bibinfo
  {journal} {New J. Phys.}\ }\textbf {\bibinfo {volume} {18}},\ \bibinfo
  {pages} {043029} (\bibinfo {year} {2016})},\ \Eprint
  {http://arxiv.org/abs/1512.00990} {arXiv:1512.00990 [quant-ph]} \BibitemShut
  {NoStop}%
\bibitem [{\citenamefont {Schutzhold}\ and\ \citenamefont
  {Unruh}(2013)}]{SchuTzhold:2013fga}%
  \BibitemOpen
  \bibfield  {author} {\bibinfo {author} {\bibfnamefont {Ralf}\ \bibnamefont
  {Schutzhold}}\ and\ \bibinfo {author} {\bibfnamefont {William~G.}\
  \bibnamefont {Unruh}},\ }\bibfield  {title} {\enquote {\bibinfo {title}
  {{Cosmological particle creation in the lab?}}}\ }\bibfield  {booktitle}
  {\emph {\bibinfo {booktitle} {{Analogue Gravity Phenomenology}}},\ }\href
  {\doibase 10.1007/978-3-319-00266-8_3} {\bibfield  {journal} {\bibinfo
  {journal} {Lect. Notes Phys.}\ }\textbf {\bibinfo {volume} {870}},\ \bibinfo
  {pages} {51--61} (\bibinfo {year} {2013})},\ \Eprint
  {http://arxiv.org/abs/1203.1173} {arXiv:1203.1173 [gr-qc]} \BibitemShut
  {NoStop}%
\bibitem [{\citenamefont {Maldacena}(1999)}]{Maldacena:1997re}%
  \BibitemOpen
  \bibfield  {author} {\bibinfo {author} {\bibfnamefont {Juan~Martin}\
  \bibnamefont {Maldacena}},\ }\bibfield  {title} {\enquote {\bibinfo {title}
  {{The Large N limit of superconformal field theories and supergravity}},}\
  }\href {\doibase 10.1023/A:1026654312961, 10.4310/ATMP.1998.v2.n2.a1}
  {\bibfield  {journal} {\bibinfo  {journal} {Int. J. Theor. Phys.}\ }\textbf
  {\bibinfo {volume} {38}},\ \bibinfo {pages} {1113--1133} (\bibinfo {year}
  {1999})},\ \bibinfo {note} {[Adv. Theor. Math. Phys.2,231(1998)]},\ \Eprint
  {http://arxiv.org/abs/hep-th/9711200} {arXiv:hep-th/9711200 [hep-th]}
  \BibitemShut {NoStop}%
\bibitem [{\citenamefont {Witten}(1998)}]{Witten:1998qj}%
  \BibitemOpen
  \bibfield  {author} {\bibinfo {author} {\bibfnamefont {Edward}\ \bibnamefont
  {Witten}},\ }\bibfield  {title} {\enquote {\bibinfo {title} {{Anti-de Sitter
  space and holography}},}\ }\href {\doibase 10.4310/ATMP.1998.v2.n2.a2}
  {\bibfield  {journal} {\bibinfo  {journal} {Adv. Theor. Math. Phys.}\
  }\textbf {\bibinfo {volume} {2}},\ \bibinfo {pages} {253--291} (\bibinfo
  {year} {1998})},\ \Eprint {http://arxiv.org/abs/hep-th/9802150}
  {arXiv:hep-th/9802150 [hep-th]} \BibitemShut {NoStop}%
\bibitem [{\citenamefont {Gubser}\ \emph {et~al.}(1998)\citenamefont {Gubser},
  \citenamefont {Klebanov},\ and\ \citenamefont {Polyakov}}]{Gubser:1998bc}%
  \BibitemOpen
  \bibfield  {author} {\bibinfo {author} {\bibfnamefont {S.~S.}\ \bibnamefont
  {Gubser}}, \bibinfo {author} {\bibfnamefont {Igor~R.}\ \bibnamefont
  {Klebanov}}, \ and\ \bibinfo {author} {\bibfnamefont {Alexander~M.}\
  \bibnamefont {Polyakov}},\ }\bibfield  {title} {\enquote {\bibinfo {title}
  {{Gauge theory correlators from noncritical string theory}},}\ }\href
  {\doibase 10.1016/S0370-2693(98)00377-3} {\bibfield  {journal} {\bibinfo
  {journal} {Phys. Lett.}\ }\textbf {\bibinfo {volume} {B428}},\ \bibinfo
  {pages} {105--114} (\bibinfo {year} {1998})},\ \Eprint
  {http://arxiv.org/abs/hep-th/9802109} {arXiv:hep-th/9802109 [hep-th]}
  \BibitemShut {NoStop}%
\bibitem [{\citenamefont {Ryu}\ and\ \citenamefont
  {Takayanagi}(2006)}]{Ryu:2006bv}%
  \BibitemOpen
  \bibfield  {author} {\bibinfo {author} {\bibfnamefont {Shinsei}\ \bibnamefont
  {Ryu}}\ and\ \bibinfo {author} {\bibfnamefont {Tadashi}\ \bibnamefont
  {Takayanagi}},\ }\bibfield  {title} {\enquote {\bibinfo {title} {{Holographic
  derivation of entanglement entropy from AdS/CFT}},}\ }\href {\doibase
  10.1103/PhysRevLett.96.181602} {\bibfield  {journal} {\bibinfo  {journal}
  {Phys. Rev. Lett.}\ }\textbf {\bibinfo {volume} {96}},\ \bibinfo {pages}
  {181602} (\bibinfo {year} {2006})},\ \Eprint
  {http://arxiv.org/abs/hep-th/0603001} {arXiv:hep-th/0603001 [hep-th]}
  \BibitemShut {NoStop}%
\bibitem [{\citenamefont {Nishioka}\ \emph {et~al.}(2009)\citenamefont
  {Nishioka}, \citenamefont {Ryu},\ and\ \citenamefont
  {Takayanagi}}]{Nishioka:2009un}%
  \BibitemOpen
  \bibfield  {author} {\bibinfo {author} {\bibfnamefont {Tatsuma}\ \bibnamefont
  {Nishioka}}, \bibinfo {author} {\bibfnamefont {Shinsei}\ \bibnamefont {Ryu}},
  \ and\ \bibinfo {author} {\bibfnamefont {Tadashi}\ \bibnamefont
  {Takayanagi}},\ }\bibfield  {title} {\enquote {\bibinfo {title} {{Holographic
  Entanglement Entropy: An Overview}},}\ }\href {\doibase
  10.1088/1751-8113/42/50/504008} {\bibfield  {journal} {\bibinfo  {journal}
  {J. Phys.}\ }\textbf {\bibinfo {volume} {A42}},\ \bibinfo {pages} {504008}
  (\bibinfo {year} {2009})},\ \Eprint {http://arxiv.org/abs/0905.0932}
  {arXiv:0905.0932 [hep-th]} \BibitemShut {NoStop}%
\bibitem [{\citenamefont {Hayden}\ and\ \citenamefont
  {Preskill}(2007)}]{Hayden:2007cs}%
  \BibitemOpen
  \bibfield  {author} {\bibinfo {author} {\bibfnamefont {Patrick}\ \bibnamefont
  {Hayden}}\ and\ \bibinfo {author} {\bibfnamefont {John}\ \bibnamefont
  {Preskill}},\ }\bibfield  {title} {\enquote {\bibinfo {title} {{Black holes
  as mirrors: Quantum information in random subsystems}},}\ }\href {\doibase
  10.1088/1126-6708/2007/09/120} {\bibfield  {journal} {\bibinfo  {journal}
  {JHEP}\ }\textbf {\bibinfo {volume} {09}},\ \bibinfo {pages} {120} (\bibinfo
  {year} {2007})},\ \Eprint {http://arxiv.org/abs/0708.4025} {arXiv:0708.4025
  [hep-th]} \BibitemShut {NoStop}%
\bibitem [{\citenamefont {Shenker}\ and\ \citenamefont
  {Stanford}(2014)}]{Shenker2014}%
  \BibitemOpen
  \bibfield  {author} {\bibinfo {author} {\bibfnamefont {Stephen~H.}\
  \bibnamefont {Shenker}}\ and\ \bibinfo {author} {\bibfnamefont {Douglas}\
  \bibnamefont {Stanford}},\ }\bibfield  {title} {\enquote {\bibinfo {title}
  {Black holes and the butterfly effect},}\ }\href {\doibase
  10.1007/jhep03(2014)067} {\bibfield  {journal} {\bibinfo  {journal} {Journal
  of High Energy Physics}\ }\textbf {\bibinfo {volume} {2014}} (\bibinfo {year}
  {2014}),\ 10.1007/jhep03(2014)067}\BibitemShut {NoStop}%
\bibitem [{\citenamefont {Sekino}\ and\ \citenamefont
  {Susskind}(2008)}]{Sekino:2008he}%
  \BibitemOpen
  \bibfield  {author} {\bibinfo {author} {\bibfnamefont {Yasuhiro}\
  \bibnamefont {Sekino}}\ and\ \bibinfo {author} {\bibfnamefont {Leonard}\
  \bibnamefont {Susskind}},\ }\bibfield  {title} {\enquote {\bibinfo {title}
  {{Fast Scramblers}},}\ }\href {\doibase 10.1088/1126-6708/2008/10/065}
  {\bibfield  {journal} {\bibinfo  {journal} {JHEP}\ }\textbf {\bibinfo
  {volume} {10}},\ \bibinfo {pages} {065} (\bibinfo {year} {2008})},\ \Eprint
  {http://arxiv.org/abs/0808.2096} {arXiv:0808.2096 [hep-th]} \BibitemShut
  {NoStop}%
\bibitem [{\citenamefont {Maldacena}\ \emph {et~al.}(2016)\citenamefont
  {Maldacena}, \citenamefont {Shenker},\ and\ \citenamefont
  {Stanford}}]{Maldacena2016}%
  \BibitemOpen
  \bibfield  {author} {\bibinfo {author} {\bibfnamefont {Juan}\ \bibnamefont
  {Maldacena}}, \bibinfo {author} {\bibfnamefont {Stephen~H.}\ \bibnamefont
  {Shenker}}, \ and\ \bibinfo {author} {\bibfnamefont {Douglas}\ \bibnamefont
  {Stanford}},\ }\bibfield  {title} {\enquote {\bibinfo {title} {A bound on
  chaos},}\ }\href {\doibase 10.1007/JHEP08(2016)106} {\bibfield  {journal}
  {\bibinfo  {journal} {Journal of High Energy Physics}\ }\textbf {\bibinfo
  {volume} {2016}},\ \bibinfo {pages} {106} (\bibinfo {year}
  {2016})}\BibitemShut {NoStop}%
\bibitem [{Note1()}]{Note1}%
  \BibitemOpen
  \bibinfo {note} {This equation is singular at horizon due to $f(x_h)=0$. A
  discussion about this point can be found in our appendix~\ref
  {app1}.}\BibitemShut {Stop}%
\bibitem [{\citenamefont {Mann}\ \emph {et~al.}(1991)\citenamefont {Mann},
  \citenamefont {Morsink}, \citenamefont {Sikkema},\ and\ \citenamefont
  {Steele}}]{PhysRevD.43.3948}%
  \BibitemOpen
  \bibfield  {author} {\bibinfo {author} {\bibfnamefont {R.~B.}\ \bibnamefont
  {Mann}}, \bibinfo {author} {\bibfnamefont {S.~M.}\ \bibnamefont {Morsink}},
  \bibinfo {author} {\bibfnamefont {A.~E.}\ \bibnamefont {Sikkema}}, \ and\
  \bibinfo {author} {\bibfnamefont {T.~G.}\ \bibnamefont {Steele}},\ }\bibfield
   {title} {\enquote {\bibinfo {title} {Semiclassical gravity in 1+1
  dimensions},}\ }\href {\doibase 10.1103/PhysRevD.43.3948} {\bibfield
  {journal} {\bibinfo  {journal} {Phys. Rev. D}\ }\textbf {\bibinfo {volume}
  {43}},\ \bibinfo {pages} {3948--3957} (\bibinfo {year} {1991})}\BibitemShut
  {NoStop}%
\bibitem [{\citenamefont {Gersch}\ and\ \citenamefont
  {Knollman}(1963)}]{PhysRev.129.959}%
  \BibitemOpen
  \bibfield  {author} {\bibinfo {author} {\bibfnamefont {H.~A.}\ \bibnamefont
  {Gersch}}\ and\ \bibinfo {author} {\bibfnamefont {G.~C.}\ \bibnamefont
  {Knollman}},\ }\bibfield  {title} {\enquote {\bibinfo {title} {Quantum cell
  model for bosons},}\ }\href {\doibase 10.1103/PhysRev.129.959} {\bibfield
  {journal} {\bibinfo  {journal} {Phys. Rev.}\ }\textbf {\bibinfo {volume}
  {129}},\ \bibinfo {pages} {959--967} (\bibinfo {year} {1963})}\BibitemShut
  {NoStop}%
\bibitem [{\citenamefont {Ma}\ \emph {et~al.}(1986)\citenamefont {Ma},
  \citenamefont {Halperin},\ and\ \citenamefont {Lee}}]{PhysRevB.34.3136}%
  \BibitemOpen
  \bibfield  {author} {\bibinfo {author} {\bibfnamefont {M.}~\bibnamefont
  {Ma}}, \bibinfo {author} {\bibfnamefont {B.~I.}\ \bibnamefont {Halperin}}, \
  and\ \bibinfo {author} {\bibfnamefont {P.~A.}\ \bibnamefont {Lee}},\
  }\bibfield  {title} {\enquote {\bibinfo {title} {Strongly disordered
  superfluids: Quantum fluctuations and critical behavior},}\ }\href {\doibase
  10.1103/PhysRevB.34.3136} {\bibfield  {journal} {\bibinfo  {journal} {Phys.
  Rev. B}\ }\textbf {\bibinfo {volume} {34}},\ \bibinfo {pages} {3136--3143}
  (\bibinfo {year} {1986})}\BibitemShut {NoStop}%
\bibitem [{\citenamefont {Giamarchi}\ and\ \citenamefont
  {Schulz}(1988)}]{PhysRevB.37.325}%
  \BibitemOpen
  \bibfield  {author} {\bibinfo {author} {\bibfnamefont {T.}~\bibnamefont
  {Giamarchi}}\ and\ \bibinfo {author} {\bibfnamefont {H.~J.}\ \bibnamefont
  {Schulz}},\ }\bibfield  {title} {\enquote {\bibinfo {title} {Anderson
  localization and interactions in one-dimensional metals},}\ }\href {\doibase
  10.1103/PhysRevB.37.325} {\bibfield  {journal} {\bibinfo  {journal} {Phys.
  Rev. B}\ }\textbf {\bibinfo {volume} {37}},\ \bibinfo {pages} {325--340}
  (\bibinfo {year} {1988})}\BibitemShut {NoStop}%
\bibitem [{\citenamefont {Fisher}\ \emph {et~al.}(1989)\citenamefont {Fisher},
  \citenamefont {Weichman}, \citenamefont {Grinstein},\ and\ \citenamefont
  {Fisher}}]{PhysRevB.40.546}%
  \BibitemOpen
  \bibfield  {author} {\bibinfo {author} {\bibfnamefont {Matthew P.~A.}\
  \bibnamefont {Fisher}}, \bibinfo {author} {\bibfnamefont {Peter~B.}\
  \bibnamefont {Weichman}}, \bibinfo {author} {\bibfnamefont {G.}~\bibnamefont
  {Grinstein}}, \ and\ \bibinfo {author} {\bibfnamefont {Daniel~S.}\
  \bibnamefont {Fisher}},\ }\bibfield  {title} {\enquote {\bibinfo {title}
  {Boson localization and the superfluid-insulator transition},}\ }\href
  {\doibase 10.1103/PhysRevB.40.546} {\bibfield  {journal} {\bibinfo  {journal}
  {Phys. Rev. B}\ }\textbf {\bibinfo {volume} {40}},\ \bibinfo {pages}
  {546--570} (\bibinfo {year} {1989})}\BibitemShut {NoStop}%
\bibitem [{\citenamefont {Hensgens}\ \emph {et~al.}(2017)\citenamefont
  {Hensgens}, \citenamefont {Fujita}, \citenamefont {Janssen}, \citenamefont
  {Li}, \citenamefont {Diepen}, \citenamefont {Reichl}, \citenamefont
  {Wegscheider}, \citenamefont {Sarma},\ and\ \citenamefont
  {Vandersypen}}]{Hensgens2017}%
  \BibitemOpen
  \bibfield  {author} {\bibinfo {author} {\bibfnamefont {T.}~\bibnamefont
  {Hensgens}}, \bibinfo {author} {\bibfnamefont {T.}~\bibnamefont {Fujita}},
  \bibinfo {author} {\bibfnamefont {L.}~\bibnamefont {Janssen}}, \bibinfo
  {author} {\bibfnamefont {Xiao}\ \bibnamefont {Li}}, \bibinfo {author}
  {\bibfnamefont {C.~J.~Van}\ \bibnamefont {Diepen}}, \bibinfo {author}
  {\bibfnamefont {C.}~\bibnamefont {Reichl}}, \bibinfo {author} {\bibfnamefont
  {W.}~\bibnamefont {Wegscheider}}, \bibinfo {author} {\bibfnamefont {S.~Das}\
  \bibnamefont {Sarma}}, \ and\ \bibinfo {author} {\bibfnamefont {L.~M.~K.}\
  \bibnamefont {Vandersypen}},\ }\bibfield  {title} {\enquote {\bibinfo {title}
  {Quantum simulation of a fermi{\textendash}hubbard model using a
  semiconductor quantum dot array},}\ }\href {\doibase 10.1038/nature23022}
  {\bibfield  {journal} {\bibinfo  {journal} {Nature}\ }\textbf {\bibinfo
  {volume} {548}},\ \bibinfo {pages} {70--73} (\bibinfo {year}
  {2017})}\BibitemShut {NoStop}%
\bibitem [{\citenamefont {Tarruell}\ and\ \citenamefont
  {Sanchez-Palencia}(2018)}]{Tarruell2018}%
  \BibitemOpen
  \bibfield  {author} {\bibinfo {author} {\bibfnamefont {Leticia}\ \bibnamefont
  {Tarruell}}\ and\ \bibinfo {author} {\bibfnamefont {Laurent}\ \bibnamefont
  {Sanchez-Palencia}},\ }\bibfield  {title} {\enquote {\bibinfo {title}
  {Quantum simulation of the hubbard model with ultracold fermions in optical
  lattices},}\ }\href {\doibase 10.1016/j.crhy.2018.10.013} {\bibfield
  {journal} {\bibinfo  {journal} {Comptes Rendus Physique}\ }\textbf {\bibinfo
  {volume} {19}},\ \bibinfo {pages} {365--393} (\bibinfo {year}
  {2018})}\BibitemShut {NoStop}%
\bibitem [{\citenamefont {Salfi}\ \emph {et~al.}(2016)\citenamefont {Salfi},
  \citenamefont {Mol}, \citenamefont {Rahman}, \citenamefont {Klimeck},
  \citenamefont {Simmons}, \citenamefont {Hollenberg},\ and\ \citenamefont
  {Rogge}}]{Salfi2016}%
  \BibitemOpen
  \bibfield  {author} {\bibinfo {author} {\bibfnamefont {J.}~\bibnamefont
  {Salfi}}, \bibinfo {author} {\bibfnamefont {J.~A.}\ \bibnamefont {Mol}},
  \bibinfo {author} {\bibfnamefont {R.}~\bibnamefont {Rahman}}, \bibinfo
  {author} {\bibfnamefont {G.}~\bibnamefont {Klimeck}}, \bibinfo {author}
  {\bibfnamefont {M.~Y.}\ \bibnamefont {Simmons}}, \bibinfo {author}
  {\bibfnamefont {L.~C.~L.}\ \bibnamefont {Hollenberg}}, \ and\ \bibinfo
  {author} {\bibfnamefont {S.}~\bibnamefont {Rogge}},\ }\bibfield  {title}
  {\enquote {\bibinfo {title} {Quantum simulation of the hubbard model with
  dopant atoms in silicon},}\ }\href {\doibase 10.1038/ncomms11342} {\bibfield
  {journal} {\bibinfo  {journal} {Nature Communications}\ }\textbf {\bibinfo
  {volume} {7}} (\bibinfo {year} {2016}),\ 10.1038/ncomms11342}\BibitemShut
  {NoStop}%
\bibitem [{\citenamefont {Barouch}\ \emph {et~al.}(1970)\citenamefont
  {Barouch}, \citenamefont {McCoy},\ and\ \citenamefont
  {Dresden}}]{PhysRevA.2.1075}%
  \BibitemOpen
  \bibfield  {author} {\bibinfo {author} {\bibfnamefont {Eytan}\ \bibnamefont
  {Barouch}}, \bibinfo {author} {\bibfnamefont {Barry~M.}\ \bibnamefont
  {McCoy}}, \ and\ \bibinfo {author} {\bibfnamefont {Max}\ \bibnamefont
  {Dresden}},\ }\bibfield  {title} {\enquote {\bibinfo {title} {Statistical
  mechanics of the $\mathrm{XY}$ model. i},}\ }\href {\doibase
  10.1103/PhysRevA.2.1075} {\bibfield  {journal} {\bibinfo  {journal} {Phys.
  Rev. A}\ }\textbf {\bibinfo {volume} {2}},\ \bibinfo {pages} {1075--1092}
  (\bibinfo {year} {1970})}\BibitemShut {NoStop}%
\bibitem [{\citenamefont {Barouch}\ and\ \citenamefont
  {McCoy}(1971)}]{PhysRevA.3.786}%
  \BibitemOpen
  \bibfield  {author} {\bibinfo {author} {\bibfnamefont {Eytan}\ \bibnamefont
  {Barouch}}\ and\ \bibinfo {author} {\bibfnamefont {Barry~M.}\ \bibnamefont
  {McCoy}},\ }\bibfield  {title} {\enquote {\bibinfo {title} {Statistical
  mechanics of the $xy$ model. ii. spin-correlation functions},}\ }\href
  {\doibase 10.1103/PhysRevA.3.786} {\bibfield  {journal} {\bibinfo  {journal}
  {Phys. Rev. A}\ }\textbf {\bibinfo {volume} {3}},\ \bibinfo {pages}
  {786--804} (\bibinfo {year} {1971})}\BibitemShut {NoStop}%
\bibitem [{\citenamefont {D.}\ and\ \citenamefont {I.}(2004)}]{PRL93-263602}%
  \BibitemOpen
  \bibfield  {author} {\bibinfo {author} {\bibfnamefont {Porras}\ \bibnamefont
  {D.}}\ and\ \bibinfo {author} {\bibfnamefont {Cirac~J.}\ \bibnamefont {I.}},\
  }\bibfield  {title} {\enquote {\bibinfo {title} {Bose-einstein condensation
  and strong-correlation behavior of phonons in ion traps},}\ }\href@noop {}
  {\bibfield  {journal} {\bibinfo  {journal} {Phys. Rev. Lett.}\ }\textbf
  {\bibinfo {volume} {93}},\ \bibinfo {pages} {263602} (\bibinfo {year}
  {2004})}\BibitemShut {NoStop}%
\bibitem [{\citenamefont {Damour}\ and\ \citenamefont
  {Ruffini}(1976)}]{Damour:1976jd}%
  \BibitemOpen
  \bibfield  {author} {\bibinfo {author} {\bibfnamefont {T.}~\bibnamefont
  {Damour}}\ and\ \bibinfo {author} {\bibfnamefont {R.}~\bibnamefont
  {Ruffini}},\ }\bibfield  {title} {\enquote {\bibinfo {title} {{Black Hole
  Evaporation in the Klein-Sauter-Heisenberg-Euler Formalism}},}\ }\href
  {\doibase 10.1103/PhysRevD.14.332} {\bibfield  {journal} {\bibinfo  {journal}
  {Phys. Rev.}\ }\textbf {\bibinfo {volume} {D14}},\ \bibinfo {pages}
  {332--334} (\bibinfo {year} {1976})}\BibitemShut {NoStop}%
\bibitem [{\citenamefont {Parikh}\ and\ \citenamefont
  {Wilczek}(2000)}]{Parikh:1999mf}%
  \BibitemOpen
  \bibfield  {author} {\bibinfo {author} {\bibfnamefont {Maulik~K.}\
  \bibnamefont {Parikh}}\ and\ \bibinfo {author} {\bibfnamefont {Frank}\
  \bibnamefont {Wilczek}},\ }\bibfield  {title} {\enquote {\bibinfo {title}
  {{Hawking radiation as tunneling}},}\ }\href {\doibase
  10.1103/PhysRevLett.85.5042} {\bibfield  {journal} {\bibinfo  {journal}
  {Phys. Rev. Lett.}\ }\textbf {\bibinfo {volume} {85}},\ \bibinfo {pages}
  {5042--5045} (\bibinfo {year} {2000})},\ \Eprint
  {http://arxiv.org/abs/hep-th/9907001} {arXiv:hep-th/9907001 [hep-th]}
  \BibitemShut {NoStop}%
\bibitem [{\citenamefont {Arzano}\ \emph {et~al.}(2005)\citenamefont {Arzano},
  \citenamefont {Medved},\ and\ \citenamefont {Vagenas}}]{Arzano:2005rs}%
  \BibitemOpen
  \bibfield  {author} {\bibinfo {author} {\bibfnamefont {Michele}\ \bibnamefont
  {Arzano}}, \bibinfo {author} {\bibfnamefont {A.~J.~M.}\ \bibnamefont
  {Medved}}, \ and\ \bibinfo {author} {\bibfnamefont {Elias~C.}\ \bibnamefont
  {Vagenas}},\ }\bibfield  {title} {\enquote {\bibinfo {title} {{Hawking
  radiation as tunneling through the quantum horizon}},}\ }\href {\doibase
  10.1088/1126-6708/2005/09/037} {\bibfield  {journal} {\bibinfo  {journal}
  {JHEP}\ }\textbf {\bibinfo {volume} {09}},\ \bibinfo {pages} {037} (\bibinfo
  {year} {2005})},\ \Eprint {http://arxiv.org/abs/hep-th/0505266}
  {arXiv:hep-th/0505266 [hep-th]} \BibitemShut {NoStop}%
\bibitem [{\citenamefont {Guhr}\ \emph {et~al.}(1998)\citenamefont {Guhr},
  \citenamefont {M¨¹ller¨CGroeling},\ and\ \citenamefont
  {Weidenm¨¹ller}}]{GUHR1998189}%
  \BibitemOpen
  \bibfield  {author} {\bibinfo {author} {\bibfnamefont {Thomas}\ \bibnamefont
  {Guhr}}, \bibinfo {author} {\bibfnamefont {Axel}\ \bibnamefont
  {M¨¹ller¨CGroeling}}, \ and\ \bibinfo {author} {\bibfnamefont {Hans~A.}\
  \bibnamefont {Weidenm¨¹ller}},\ }\bibfield  {title} {\enquote {\bibinfo
  {title} {Random-matrix theories in quantum physics: common concepts},}\
  }\href {\doibase https://doi.org/10.1016/S0370-1573(97)00088-4} {\bibfield
  {journal} {\bibinfo  {journal} {Physics Reports}\ }\textbf {\bibinfo {volume}
  {299}},\ \bibinfo {pages} {189 -- 425} (\bibinfo {year} {1998})}\BibitemShut
  {NoStop}%
\bibitem [{\citenamefont {Jafarizadeh}\ \emph {et~al.}(2012)\citenamefont
  {Jafarizadeh}, \citenamefont {Fouladi}, \citenamefont {Sabri},\ and\
  \citenamefont {Maleki}}]{Jafarizadeh:2012da}%
  \BibitemOpen
  \bibfield  {author} {\bibinfo {author} {\bibfnamefont {M.~A.}\ \bibnamefont
  {Jafarizadeh}}, \bibinfo {author} {\bibfnamefont {N.}~\bibnamefont
  {Fouladi}}, \bibinfo {author} {\bibfnamefont {H.}~\bibnamefont {Sabri}}, \
  and\ \bibinfo {author} {\bibfnamefont {B.~R.}\ \bibnamefont {Maleki}},\
  }\bibfield  {title} {\enquote {\bibinfo {title} {{Investigation of Level
  Statistics by Generalized Brody Distribution and Maximum Likelihood
  Estimation Method}},}\ }\href@noop {} {\  (\bibinfo {year} {2012})},\ \Eprint
  {http://arxiv.org/abs/1210.4751} {arXiv:1210.4751 [nucl-th]} \BibitemShut
  {NoStop}%
\end{thebibliography}%
\end{document}